\pdfoutput=1
\documentclass[floatfix,showpacs,amsmath,amssymb,twocolumn,pra,longbibliography,superscriptaddress,notitlepage]{revtex4-1}
\usepackage[T1]{fontenc}
\usepackage{comment}
\usepackage[utf8]{inputenc}
\usepackage{lmodern}
\usepackage{amsmath,amssymb,amsthm,mathrsfs,amsfonts,dsfont}
\usepackage{color}
\usepackage{graphicx}
\usepackage{braket}
\usepackage{bm}
\usepackage[svgnames,psnames]{xcolor}
\usepackage[colorlinks,citecolor=DarkGreen,linkcolor=FireBrick,urlcolor=FireBrick,linktocpage,unicode]{hyperref}
\usepackage[whole]{bxcjkjatype}

\allowdisplaybreaks[3]

\begin{document}

\title{Possibility of the total thermodynamic entropy production rate of a finite-sized isolated quantum system to be negative for the Gorini-Kossakowski-Sudarshan-Lindblad-type Markovian dynamics of its subsystem
}

\author{Takaaki Aoki}
\email{t-aoki@iis.u-tokyo.ac.jp, takaaki-aoki@aist.go.jp}
\affiliation{Department of Physics, The University of Tokyo, 5-1-5 Kashiwanoha, Kashiwa, Chiba 277-8574, Japan}
\affiliation{Research Center for Emerging Computing Technologies (RCECT), National Institute of Advanced Industrial Science and Technology (AIST), 1-1-1 Umezono, Tsukuba, Ibaraki 305-8568, Japan}

\author{Yuichiro Matsuzaki}
\email{matsuzaki.yuichiro@aist.go.jp}
\affiliation{Research Center for Emerging Computing Technologies (RCECT), National Institute of Advanced Industrial Science and Technology (AIST), 1-1-1 Umezono, Tsukuba, Ibaraki 305-8568, Japan}

\author{Hideaki Hakoshima}
\email{hakoshima-hideaki@aist.go.jp}
\affiliation{Research Center for Emerging Computing Technologies (RCECT), National Institute of Advanced Industrial Science and Technology (AIST), 1-1-1 Umezono, Tsukuba, Ibaraki 305-8568, Japan}

\begin{abstract}
We investigate a total thermodynamic entropy production
rate of an isolated quantum system.
In particular,
we consider a quantum model of coupled harmonic oscillators in a star configuration,
where a central harmonic oscillator
(system) is coupled to a finite number of surrounding harmonic oscillators (bath).
In this model, when the initial state of the total system is given by the tensor product of the
Gibbs states of the system and the bath,
every harmonic oscillator is always in a Gibbs state with a time-dependent temperature.
This enables us to define time-dependent thermodynamic entropy for each harmonic oscillator and total nonequilibrium thermodynamic
entropy as the summation of them.
We analytically
confirm that the total thermodynamic entropy
satisfies the third law of thermodynamics.
Our numerical solutions show that,
even when the dynamics of the system is
well approximated by the
Gorini-Kossakowski-Sudarshan-Lindblad (GKSL)-type
Markovian master equation,
the total thermodynamic entropy production rate
can be negative, while the total thermodynamic entropy satisfies
the second law of thermodynamics.
This result is a counterexample to the common belief
that the total entropy production rate is non-negative
when the system is under
the GKSL-type
Markovian dynamics.
\end{abstract}
\maketitle

\section{Introduction}
Thermodynamics of macroscopic systems 
explains their macroscopic thermodynamic changes,
where microscopic fluctuations can be neglected
\cite{oono2017perspectives,callen1985thermodynamics,lebon2008understanding}.
On the other hand,
because of the development of nanotechnology,
researchers in recent years have tried
to extend the conventional thermodynamics
to the microscopic world,
where not only thermal but also quantum
fluctuations cannot be neglected.
This research field is called
quantum thermodynamics
\cite{binder2018thermodynamics,doi:10.1080/00107514.2016.1201896,e15062100}
which explains microscopic thermodynamic changes
of microscopic systems and macroscopic ones.

One of the most fundamental problems in quantum thermodynamics
is how to define thermodynamic quantities such as thermodynamic
entropy, temperature, heat and work.
Of particular importance is the
thermodynamic entropy
because it characterizes the irreversibility of thermodynamics.
This is why researchers have
suggested several definitions of thermodynamic entropy
\cite{strasberg2021second,PhysRevA.99.012103,doi:10.1142/9789811211720_0014}
and various ones of
entropy production and
of its rate; see Ref.~\cite{landi2020irreversible} and
references therein.
However, there is no consensus for now.
The connection between the entropy production and
quantum information \cite[Sec.~5.4]{Goold_2016}
such as quantum cryptography \cite{tan2020computing}
is also currently being investigated.

One of the typical setups in quantum thermodynamics is
a quantum system 
coupled with a bath.
The system is open 
\cite{rivas2012open,breuer2002theory}
and the total system,
which is a compound of the system and the bath,
is isolated (closed)
when the total Hamiltonian is time-independent
(time-dependent)
\cite[Sec.~3.1.1]{breuer2002theory}.
There is active research
\cite{marcantoni2017entropy,PhysRevA.95.012122,PhysRevA.98.012130,PhysRevE.98.032102,PhysRevE.99.012120,PhysRevLett.124.160601}
into
the relation between non-Markovianity \cite{RevModPhys.89.015001} of the dynamics of an open quantum system and a negative entropy production rate of
the total system.
However, there is no agreement
about this relation
mainly because there is no unified definition of the entropy production rate
or of non-Markovianity.
On the other hand,
when an open quantum
system is under the 
Gorini-Kossakowski-Sudarshan-Lindblad (GKSL)-type
Markovian dynamics \cite{breuer2002theory,rivas2012open,Rivas_2010,doi:10.1063/1.522979,lindblad1976generators},
it is widely believed that the
entropy production rate of the total system
is non-negative
\cite{marcantoni2017entropy,PhysRevA.95.012122,PhysRevA.98.012130,PhysRevE.98.032102,PhysRevE.99.012120,PhysRevLett.124.160601},
and researchers often use the von Neumann
entropy production rate \cite{doi:10.1063/1.523789},
which is the minus time-derivative of the von Neumann
relative entropy \cite[Sec.~11.8]{wilde2013quantum}
between the reduced state of the system and the reference stationary state of the GKSL master equation.
As we will see later, there is 
an implicit assumption in the form of the von Neumann entropy
production rate that the size of a bath
is so macroscopically large that its temperature
does not change during the dynamics.
However, when the size of the bath is finite,
the temperature of the bath varies with time in general
\cite{strasberg2021second}.
Then, we cannot use the von Neumann entropy production rate.

In this paper,
we define and investigate a total thermodynamic entropy production rate of an isolated quantum system
which consists of a system and a finite-sized bath. 
In particular,
we consider
a quantum model of coupled harmonic oscillators in
a star configuration.
We show that, contrary to the common belief,
the entropy production rate of
the total system can be negative
even when the dynamics of the central harmonic oscillator
(system) is well approximated by the GKSL master equation.
This comes from
the temperature-changes of the surrounding
harmonic oscillators (bath).

This paper is organized as follows.
In Sec.~\ref{sec:review},
we review thermodynamic entropy
of macroscopic systems
and the von Neumann entropy production rate.
In Sec.~\ref{sec:Settings},
we introduce
our model, the initial state, and the dynamics.
In Sec.~\ref{sec:results1},
we show that
every harmonic oscillator is in
a Gibbs state with a time-dependent temperature
in our settings.
We thus define the time-dependent thermodynamic entropy
of each harmonic oscillator in a similar way to
the definition in equilibrium thermodynamics
and statistical mechanics.
Then, we define the non-equilibrium thermodynamic
entropy of the total
system as the summation of them.
This total thermodynamic entropy satisfies the third law
of thermodynamics.
In Sec.~\ref{sec:results2},
considering the GKSL-type Markovian dynamics
of the system,
we show numerically that our total thermodynamic
entropy production rate can take negative values,
while our total thermodynamic
entropy satisfies the second law of
thermodynamics.
In Sec.~\ref{sec:conclusion},
we draw a conclusion.

\section{Review of thermodynamic entropy and of entropy
production rate}
\label{sec:review}
\subsection{Thermodynamic entropy of macroscopic systems}
\label{subsec:macro}
Equilibrium thermodynamics of macroscopic
systems is an established theory
\cite{oono2017perspectives,callen1985thermodynamics}.
The irreversibility of thermodynamics is expressed by 
its second law,
which can be cast into the form of the
principle of increasing total thermodynamic entropy \cite[Sec.~14.2]{oono2017perspectives}
\footnote{Throughout the present paper,
we intentionally use the term
thermodynamic entropy to
distinguish it from other types of entropy,
such as von Neumann entropy \cite{von2018mathematical},
R{\'e}nyi entropy \cite{renyi1961measures},
and a diagonal entropy \cite{POLKOVNIKOV2011486}}.
Let us prepare an adiabatic system
in an equilibrium state
with some constraints (for example, a system
consisting of the two
subsystems with different temperatures
separated by an adiabatic wall).
If we get rid of the constraints
(e.g. remove the wall)
at time $t_{\mathrm{ini}}$,
the system would change to a new
equilibrium state at time $t_{\mathrm{fin}}$.
The final total thermodynamic entropy
$S^{\mathrm{th}}_{\mathrm{tot}}(t_{\mathrm{fin}})$
must be greater than or equal to the initial one $S^{\mathrm{th}}_{\mathrm{tot}}(t_{\mathrm{ini}})$:
\begin{align}
  \Delta S^{\mathrm{th}}_{\mathrm{tot}}(t_{\mathrm{fin}})
  =S^{\mathrm{th}}_{\mathrm{tot}}(t_{\mathrm{fin}})
  -S^{\mathrm{th}}_{\mathrm{tot}}(t_{\mathrm{ini}})\geq0,
  \label{eq:2nd_1}
\end{align}
where $\Delta S^{\mathrm{th}}_{\mathrm{tot}}(t)
:=S^{\mathrm{th}}_{\mathrm{tot}}(t)
-S^{\mathrm{th}}_{\mathrm{tot}}(t_{\mathrm{ini}})$
denotes the total thermodynamic entropy production
from $t_{\mathrm{ini}}$ to $t$.
This is the principle of increasing total thermodynamic
entropy.
Here, the word ``total'' refers to the adiabatic system
itself, excluding its environment,
and is used to distinguish $\Delta S^{\mathrm{th}}_{\mathrm{tot}}(t)$ from the internal
thermodynamic entropy production,
which we will explain later.
Note that the principle deals with the 
thermodynamic-entropy difference only
between the initial and final 
equilibrium states.
This does not forbid the total
thermodynamic entropy from decreasing
during the intermediate nonequilibrium process
\cite[Sec.~14.2]{oono2017perspectives}.
In other words, 
the total thermodynamic entropy
production rate $\Pi^{\mathrm{th}}_{\mathrm{tot}}(t)
:=\mathrm{d}S^{\mathrm{th}}_{\mathrm{tot}}(t)/\mathrm{d}t$
can be negative for some time $t$.

Actually, the theory of nonequilibrium thermodynamics
of macroscopic systems, including a proper definition
of non-equilibrium thermodynamic entropy
$S^{\mathrm{th}}(t)$,
has not been established yet
\cite{lebon2008understanding}.
However, the entropy balance
\cite[Sec.~2.3]{lebon2008understanding},
which we will explain below, is considered to hold
universally.
Let us consider a system $A$ and its environment $B$,
whose thermodynamic entropies
are defined as
$S^{\mathrm{th}}_A(t)$ and $S^{\mathrm{th}}_B(t)$,
respectively.
The time derivative of $S^{\mathrm{th}}_A(t)$
is written as the sum of the internal
thermodynamic entropy production rate of the system
$\mathrm{d}^{\mathrm{int}}S^{\mathrm{th}}_A(t)
/\mathrm{d}t$ and the thermodynamic
entropy flux into the system
$\mathrm{d}^{\mathrm{ext}}S^{\mathrm{th}}_A(t)
/\mathrm{d}t$ as follows:
\begin{align}
    \frac{\mathrm{d}}{\mathrm{d}t}
    S^{\mathrm{th}}_A(t)
    =\frac{\mathrm{d}^{\mathrm{int}}}{\mathrm{d}t}
    S^{\mathrm{th}}_A(t)
    +\frac{\mathrm{d}^{\mathrm{ext}}}{\mathrm{d}t}
    S^{\mathrm{th}}_A(t).
    \label{eq:SbalanceA1}
\end{align}
This is the entropy balance.
We must distinguish the internal thermodynamic entropy
production, which is the time-integral of its rate,
from the total one.
When the temperature $T_A(t)$ of the system $A$
is defined, the entropy flux into the system $A$
is defined as \cite[Sec.~1.3.3.2]{lebon2008understanding}
\begin{align}
    \frac{\mathrm{d}^{\mathrm{ext}}}{\mathrm{d}t}
    S^{\mathrm{th}}_A(t)
    :=\frac{1}{T_A(t)}
    \frac{\dj Q_A(t)}{\mathrm{d}t},
    \label{eq:extS1}
\end{align}
where $\dj Q_A(t)/\mathrm{d}t$ is the heat flux
into the system $A$ and the bar in $\dj Q_A(t)$
means that it is an inexact differential
\cite[Sec.~1.3.2]{lebon2008understanding}.
Then, the internal entropy production rate
of the system $A$ is
determined from Eqs.~\eqref{eq:SbalanceA1} and
\eqref{eq:extS1}.
On the other hand, when the temperature $T_A(t)$ of the system $A$ is not defined, it is a subject of
research how to define
$\mathrm{d}^{\mathrm{int}}S^{\mathrm{th}}_A(t)
/\mathrm{d}t$
and
$\mathrm{d}^{\mathrm{ext}}S^{\mathrm{th}}_A(t)
/\mathrm{d}t$.
A similar relation to Eq.~\eqref{eq:SbalanceA1}
holds for the environment:
\begin{align}
    \frac{\mathrm{d}}{\mathrm{d}t}
    S^{\mathrm{th}}_B(t)
    =\frac{\mathrm{d}^{\mathrm{int}}}{\mathrm{d}t}
    S^{\mathrm{th}}_B(t)
    +\frac{\mathrm{d}^{\mathrm{ext}}}{\mathrm{d}t}
    S^{\mathrm{th}}_B(t).
    \label{eq:SbalanceB1}
\end{align}
The point is that the entropy flux into the system
does not equal that out of the
environment in general:
\begin{align}
    \frac{\mathrm{d}^{\mathrm{ext}}}{\mathrm{d}t}
    S^{\mathrm{th}}_A(t)
    \neq
    -\frac{\mathrm{d}^{\mathrm{ext}}}{\mathrm{d}t}
    S^{\mathrm{th}}_B(t).
\end{align}

In order to recognize this point,
let us consider the following example
\cite[Sec.~4.3]{callen1985thermodynamics}
\cite[Sec.~7.1.1]{lebon2008understanding}.
Prepare an isolated
system composed of the two subsystems $A$ and $B$.
There are two fixed walls between $A$ and $B$:
an adiabatic wall and a diathermal wall.
The subsystem $A$ ($B$) is in an equilibrium state
with temperature $T_A^0$ ($T_B^0(>T_A^0$)).
The total system is also at equilibrium.
Then, remove the adiabatic wall at time $t_{\mathrm{ini}}$,
and heat begins to flow from $B$ to $A$
through the diathermal wall
and continues flowing until the two subsystems
are of equal temperature at time $t_{\mathrm{fin}}$.
Let us assume that
the thermal conductivity of the diathermal wall is
so small that each of the two subsystems
should be always in
an equilibrium state and that the temperatures of
them $T_A(t)$ and $T_B(t)$
change very slowly during the process.
We call this process as quasistatic
\cite[Sec.~4.3]{callen1985thermodynamics}
for both $A$ and $B$ in the meaning that
they are always in an equilibrium state.
We note that there are other definitions
of quasistatic processes; see, for example,
Sec.~12.6 in Ref.~\cite{oono2017perspectives}.

Let us describe the internal energy
of the subsystem $A$ as $E_A(t)$.
From the first law of thermodynamics,
the change of $E_A(t)$ equals to
the sum of the heat $Q_A$ into $A$
and the work $W_A$ done on $A$:
$\Delta E_A(t)=Q_A+W_A$.
In the present example,
$W_A$ is always zero because of the fixed diathermal
wall.
Hence the heat flux into the subsystem $A$ is given by
$\mathrm{d}E_A(t)/\mathrm{d}t$.
From the law of energy conservation,
the heat flux into the subsystem $B$
is given by
$-\mathrm{d}E_A(t)/\mathrm{d}t$.
Then the entropy fluxes into the two subsystems
are defined as
\begin{align}
    \frac{\mathrm{d}^{\mathrm{ext}}}{\mathrm{d}t}
    S^{\mathrm{th}}_A(t)
    &=\frac{1}{T_A(t)}
    \frac{\mathrm{d}E_A(t)}{\mathrm{d}t},
    \label{eq:efluxA1} \\
    \frac{\mathrm{d}^{\mathrm{ext}}}{\mathrm{d}t}
    S^{\mathrm{th}}_B(t)
    &=-\frac{1}{T_B(t)}
    \frac{\mathrm{d}E_A(t)}{\mathrm{d}t}.
\end{align}
These equations show
that the entropy flux into $A$
does not equal that out of $B$ at $t\neq t_{\mathrm{fin}}$
because $T_A(t)\neq T_B(t)$.
As the process is quasistatic for both $A$ and $B$,
the time derivatives of the thermodynamic entropies of the 
two subsystems are given by
\cite[Sec.~4.3]{callen1985thermodynamics}
\begin{align}
    \frac{\mathrm{d}}{\mathrm{d}t}
    S^{\mathrm{th}}_A(t)
    &=\frac{1}{T_A(t)}
    \frac{\mathrm{d}E_A(t)}{\mathrm{d}t},
    \label{eq:dSAdt1}\\
    \frac{\mathrm{d}}{\mathrm{d}t}
    S^{\mathrm{th}}_B(t)
    &=-\frac{1}{T_B(t)}
    \frac{\mathrm{d}E_A(t)}{\mathrm{d}t}.
    \label{eq:dSBdt1}
\end{align}
Combining Eqs.~\eqref{eq:SbalanceA1},
\eqref{eq:SbalanceB1}, and
\eqref{eq:efluxA1}-\eqref{eq:dSBdt1},
we find that 
the internal thermodynamic entropy
production rates of the two subsystems are both zero:
\begin{align}
    \frac{\mathrm{d}^{\mathrm{int}}}{\mathrm{d}t}
    S^{\mathrm{th}}_A(t)
    =\frac{\mathrm{d}^{\mathrm{int}}}{\mathrm{d}t}
    S^{\mathrm{th}}_B(t)
    =0.
\end{align}
We regard this
as the sign that the process is quasi-static
for both $A$ and $B$.

Let us confirm
that the above example satisfies the principle
of increasing
total thermodynamic entropy
\eqref{eq:2nd_1}.
The total thermodynamic entropy production rate
is the sum of the variation rates of the thermodynamic
entropies of the two subsystems:
\begin{align}
    \Pi^{\mathrm{th}}_{\mathrm{tot}}(t)
    &=\frac{\mathrm{d}}{\mathrm{d}t}
    S^{\mathrm{th}}_{\mathrm{tot}}(t)
    =\frac{\mathrm{d}}{\mathrm{d}t}
    S^{\mathrm{th}}_A(t)
    +\frac{\mathrm{d}}{\mathrm{d}t}
    S^{\mathrm{th}}_B(t)
    \notag \\
    &=\frac{T_B(t)-T_A(t)}{T_A(t)T_B(t)}
    \frac{\mathrm{d}E_A(t)}{\mathrm{d}t}
    \geq0,
    \label{eq:ex2nd1}
\end{align}
where the last inequality follows
from $T_B(t)\geq T_A(t)$ and
$\mathrm{d}E_A(t)/\mathrm{d}t\geq0$.
This leads to the satisfaction of the principle
of increasing total thermodynamic entropy:
\begin{align}
    \Delta S^{\mathrm{th}}_{\mathrm{tot}}
    (t_{\mathrm{fin}})
    =\int_{t_{\mathrm{ini}}}^{t_{\mathrm{fin}}}
    \mathrm{d}t\,\Pi^{\mathrm{th}}_{\mathrm{tot}}(t)
    \geq0.
\end{align}
Note that the total thermodynamic entropy production rate
is not the sum of the internal
thermodynamic entropy production rates:
\begin{align}
    \Pi^{\mathrm{th}}_{\mathrm{tot}}(t)
    \neq
    \frac{\mathrm{d}^{\mathrm{int}}}{\mathrm{d}t}
    S^{\mathrm{th}}_A(t)
    +\frac{\mathrm{d}^{\mathrm{int}}}{\mathrm{d}t}
    S^{\mathrm{th}}_B(t).
\end{align}

\subsection{The von Neumann entropy production rate}
\label{subsec:von}
Let us consider
an undriven open quantum
system $A$ (different from $A$ in the previous section)
which is coupled to a thermal bath $B$ with
initial temperature $T_B^0$.
If the coupling is
sufficiently weak, the dynamics of the system is well approximated by the GKSL-type Markovian master equation.
Then the following von Neumann entropy production rate
\cite{doi:10.1063/1.523789}
is typically used:
\begin{align}
  \Pi^{\mathrm{vN}}(t):=-\frac{\mathrm{d}}{\mathrm{d}t}
  K^{\mathrm{vN}}\left(
  \hat{\rho}_A(t)||\hat{\rho}_A^{\mathrm{th}}
  \right),
  \label{eq:vonEPR1}
\end{align}
where
$\hat{\rho}_A(t)$ is the density operator
of the system,
$\hat{\rho}_{A}^{\mathrm{th}}=
\mathrm{e}^{-\beta_B^0\hat{H}_A}
/\mathrm{Tr}[\mathrm{e}^{-\beta_B^0\hat{H}_A}]$
with $\beta_B^0=1/(k_BT_B^0)$
and with $\hat{H}_A$ being the Hamiltonian of the system
is the steady state
of the GKSL equation,
and
\begin{align}
  K^{\mathrm{vN}}
  \left(\hat{\rho}_1||\hat{\rho}_2\right)
  &:=k_B\mathrm{Tr}\left[\hat{\rho}_1
  \left(\ln\hat{\rho}_1-\ln\hat{\rho}_2\right)\right]
  \notag \\
  &=-S^{\mathrm{vN}}(\hat{\rho}_1)
  -k_B\mathrm{Tr}\left[\hat{\rho}_1\ln\hat{\rho}_2
  \right]
\end{align}
is the von Neumann relative entropy
\cite[Sec.~11.8]{wilde2013quantum}
with $S^{\mathrm{vN}}(\hat{\rho}):=
-k_B\mathrm{Tr}\left[\hat{\rho}\ln\hat{\rho}\right]$
being the von Neumann entropy.

We can transform Eq.~\eqref{eq:vonEPR1} as follows
\cite[Sec.~3.2.5]{breuer2002theory}:
\begin{align}
  \Pi^{\mathrm{vN}}(t)&=-\frac{\mathrm{d}}{\mathrm{d}t}
  K^{\mathrm{vN}}\left(
  \hat{\rho}_A(t)||\hat{\rho}_A^{\mathrm{th}}
  \right)
  \notag \\
  &=\frac{\mathrm{d}}{\mathrm{d}t}
  S^{\mathrm{vN}}_A(t)
  +k_B\frac{\mathrm{d}}{\mathrm{d}t}
  \mathrm{Tr}\left[
  \hat{\rho}_A(t)
  \ln\frac{\mathrm{e}^{-\beta_B^0\hat{H}_A}}
  {\mathrm{Tr}[\mathrm{e}^{-\beta_B^0\hat{H}_A}]}
  \right]
  \notag \\
  &=\frac{\mathrm{d}}{\mathrm{d}t}
  S^{\mathrm{vN}}_A(t)
  -\frac{1}{T_B^0}
  \frac{\mathrm{d}}{\mathrm{d}t}
  \mathrm{Tr}\left[\hat{\rho}_A(t)\hat{H}_A\right]
  \notag \\
  &\quad\mbox{}-k_B
  \ln \mathrm{Tr}[\mathrm{e}^{-\beta_B^0\hat{H}_A}]
  \frac{\mathrm{d}}{\mathrm{d}t}
  \mathrm{Tr}\left[\hat{\rho}_A(t)\right]
  \notag \\
  &=\frac{\mathrm{d}}{\mathrm{d}t}
  S^{\mathrm{vN}}_A(t)
  -\frac{1}{T_B^0}\frac{\mathrm{d}}{\mathrm{d}t}E_A(t)
  \notag \\
  &=\frac{\mathrm{d}}{\mathrm{d}t}
  S^{\mathrm{vN}}_A(t)
  +\frac{1}{T_B^0}\frac{\mathrm{d}}{\mathrm{d}t}E_B(t),
  \label{eq:vonEPR2}
\end{align}
where $E_A(t)$ ($E_B(t)$) is the mean energy of
the system (bath).
The last term in the third line of
Eq.~\eqref{eq:vonEPR2} becomes zero because
$\mathrm{Tr}\left[\hat{\rho}_A(t)\right]=1$
all the time.
From the conservation of the total energy, we
have derived
the last line
in Eq.~\eqref{eq:vonEPR2},
ignoring the interaction energy
due to weak coupling.
The first term in the last line of
Eq.~\eqref{eq:vonEPR2} is the time derivative of the von Neumann
entropy of the system and
the second term is the time-derivative of
the thermodynamic entropy
of the bath under the quasistatic process.
Note that there
appears an implicit assumption
that the temperature
of the bath does not change from the initial temperature
$T_B^0$ in this second term.
However, when the size of the bath is finite,
the temperature of a part of the bath changes as we will show
in Sec.~\ref{subsec:tworates}.
Then, we cannot use the von Neumann entropy production rate.

If we regarded the von Neumann entropy of the system
as its non-equilibrium thermodynamic entropy, the von Neumann
entropy production rate \eqref{eq:vonEPR2} would be regarded as
the total thermodynamic entropy production rate.
However, this is a delicate matter, because the von Neumann
entropy does not equal the thermodynamic entropy
in general.
For example, let us decouple the system from the bath in the middle of the dynamics.
Then the system is isolated, in general out of equilibrium,
and undergoes the unitary dynamics.
If the system shows thermalization \cite{doi:10.1080/00018732.2016.1198134},
its nonequilibrium thermodynamic entropy should change.
However, its von Neumann entropy does not change under
the unitary dynamics \cite[Sec.~11.1.1]{wilde2013quantum}.
Hence we do not regard the
von Neumann entropy
of the system as
its thermodynamic entropy in general.
However, when the system is in a Gibbs state,
its von Neumann entropy coincides with its thermodynamic
entropy.
Actually, we will consider such a case by adopting special
settings in the next section.

It is shown that the von Neumann entropy
production rate is always non-negative during the
dynamics \cite{doi:10.1063/1.523789}:
\begin{align}
  \Pi^{\mathrm{vN}}(t)\geq0\quad^{\forall}t.
  \label{eq:posiPi1}
\end{align}
This leads to the non-negative von Neumann
entropy production:
\begin{align}
  \Delta S^{\mathrm{vN}}(t):=
  \int_{t_{\mathrm{ini}}}^t\mathrm{d}s
  \,\Pi^{\mathrm{vN}}(s)
  \geq0
  \quad^{\forall}
  t\geq t_{\mathrm{ini}}.
  \label{eq:2nd_2}
\end{align}
The above two inequalities are often regarded as signs of irreversibility.
Here the total system
is not necessarily
at equilibrium at $t_{\mathrm{ini}}$ or $t$.
Hence, inequality \eqref{eq:2nd_2}
with $t=t_{\mathrm{fin}}$ is different
from the principle of increasing total thermodynamic
entropy
\eqref{eq:2nd_1}
unless each of
the total system and the system $A$
is in an equilibrium state
at both $t_{\mathrm{ini}}$ and
$t_{\mathrm{fin}}$.

\section{Settings}
\label{sec:Settings}
\subsection{Hamiltonian}
\label{subsec:model}
We consider a quantum model of
coupled harmonic oscillators
in a star configuration.
It consists of a central harmonic
oscillator $j=1$, which we refer to as system $A$, and $N$
surrounding harmonic
oscillators $j=2,\dots,N+1$, which we refer to as bath $B$.
The system $A$ and each harmonic oscillator $j$ in $B$
interact with each other with the coupling constant $g_j$.
The total system is isolated, and hence its Hamiltonian is
time-independent as in
\begin{align}
  \hat{H}=\hat{H}_A+\hat{H}_B+\hat{H}_I,
\end{align}
where
\begin{align}
  \hat{H}_A&=\hbar\omega_1\left(
  \hat{a}_1^{\dag}\hat{a}_1+\frac{1}{2}
  \right),\\
  \hat{H}_B&=\sum_{j=2}^{N+1}
  \hbar\omega_j
  \left(
  \hat{a}_j^{\dag}\hat{a}_j+\frac{1}{2}
  \right), \label{eq:HamiB1} \\
  \hat{H}_I&=\sum_{j=2}^{N+1}
  \hbar g_j
  \left(
  \hat{a}_1^{\dag}\hat{a}_j+
  \hat{a}_1\hat{a}_j^{\dag}\right),
  \label{eq:HamiI1}
\end{align}
with $\hat{a}_j$ ($\hat{a}_j^{\dag}$)
denoting the annihilation (creation) operator
of the $j$th harmonic oscillator, which satisfies the following commutation
relations:
\begin{align}
  \left[
  \hat{a}_j,\hat{a}_k^{\dag}
  \right]
  =&\delta_{j,k},
  \\
  \left[
  \hat{a}_j,\hat{a}_k
  \right]
  =&\left[
  \hat{a}_j^{\dag},\hat{a}_k^{\dag}
  \right]
  =0
  \quad\mathrm{for}\quad j,k=1,\dots,N+1.
\end{align}
This total Hamiltonian is a type of Fano-Anderson Hamiltonian
in condensed matter physics and of Lee-Friedrichs Hamiltonian in atomic physics \cite{PhysRev.109.1492,PhysRev.124.41,PhysRev.124.1866,https://doi.org/10.1002/cpa.3160010404,PhysRev.95.1329,PhysRevLett.115.168902}.
If the counter-rotating terms
$\sum_{j=2}^{N+1}\hbar g_j
(\hat{a}_1^{\dag}\hat{a}_j^{\dag}+
\hat{a}_1\hat{a}_j)$
are added to the interaction Hamiltonian
in Eq.~\eqref{eq:HamiI1},
the total Hamiltonian will become the
Caldeira-Leggett Hamiltonian
\cite{CALDEIRA1983587,PhysRevLett.115.168902}.
When $N$ is large enough, the system is
damped by the bath, and is called
a damped harmonic oscillator
\cite{Rivas_2010,rivas2012open}.
We can cast the total Hamiltonian into the form
\begin{align}
  \hat{H}&=\sum_{j=1}^{N+1}
  \frac{\hbar\omega_j}{2}
  \left(\hat{r}^2_{2j-1}
  +\hat{r}^2_{2j}
  \right)
  +\sum_{j=2}^{N+1}
  \hbar g_j
  \left(\hat{r}_1\hat{r}_{2j-1}
  +\hat{r}_2\hat{r}_{2j}
  \right) \notag\\
  &=:\frac{\hbar}{2}
  \hat{\bf{r}}^{\mathrm{T}}H
  \hat{\bf{r}},
  \label{eq:totH1}
\end{align}
where we have introduced
the modified position operator
$\hat{r}_{2j-1}$ and the modified momentum
operator $\hat{r}_{2j}$,
\begin{align}
  \hat{r}_{2j-1}:=\frac{\hat{a}_j+\hat{a}_j^{\dag}}{\sqrt{2}},
  \quad
  \hat{r}_{2j}:=\frac{\hat{a}_j-\hat{a}_j^{\dag}}
  {\sqrt{2}\,\mathrm{i}},
  \label{eq:modified}
\end{align}
and their vector representation
\begin{align}
  \hat{\bf{r}}
  =\left(
  \hat{r}_1,\hat{r}_2,\dots,
  \hat{r}_{2N+1},\hat{r}_{2N+2}
  \right)^{\mathrm{T}}
\end{align}
as well as a $2(N+1)$-dimensional symmetric matrix $H$,
whose nonzero elements are
\begin{align}
  \begin{gathered}
  H_{2j-1,2j-1}=H_{2j,2j}=\omega_j\quad\mathrm{for}\quad j=1,\dots,N+1,\\
  H_{1,2j-1}=H_{2j-1,1}=H_{2,2j}=H_{2j,2}=g_j \\
  \mathrm{for}\quad j=2,\dots,N+1.
  \end{gathered}
  \label{eq:Helements1}
\end{align}

\subsection{Initial state and unitary dynamics}
\label{subsec:initial}
Let us impose the constraint $\hat{H}_I=0$
for $t<0$
and prepare the following initial state:
\begin{align}
  \hat{\rho}(t\leq0)&=
  \frac{\mathrm{e}^{-\beta_A^0\hat{H}_A}}
  {Z_A}\otimes
  \frac{\mathrm{e}^{-\beta_B^0\hat{H}_B}}
  {Z_B} \notag \\
  &=\frac{\mathrm{e}^{-\beta_A^0\hat{H}_A}}
  {Z_A}\otimes
  \left(\bigotimes_{j=2}^{N+1}
  \frac{\mathrm{e}^{-\beta_B^0\hat{H}_j}}
  {Z_j}
  \right),
  \label{eq:ini2}
\end{align}
where
\begin{gather}
  Z_A=\mathrm{Tr}\left[
  \mathrm{e}^{-\beta_A^0\hat{H}_A}
  \right],
  \quad
  Z_B=\mathrm{Tr}
  \left[
  \mathrm{e}^{-\beta_B^0\hat{H}_B}
  \right],
  \\
  \hat{H}_j=\hbar\omega_j
  \left(
  \hat{a}_j^{\dag}\hat{a}_j+\frac{1}{2}
  \right),
  \quad
  Z_j=\mathrm{Tr}
  \left[
  \mathrm{e}^{-\beta_B^0\hat{H}_j}
  \right].
\end{gather}
That is,
the system and the bath are both in
the Gibbs states with inverse temperatures
$\beta_A^0$ and $\beta_B^0$, respectively,
and they are uncorrelated.
Because of the constraint $\hat{H}_I=0$,
the initial state \eqref{eq:ini2} is an equilibrium
state:
\begin{align}
  \hat{\rho}(t_2)&=
  e^{-\mathrm{i}\frac{\hat{H}_A+\hat{H}_B}
  {\hbar}(t_2-t_1)
  }
  \hat{\rho}(t_1)
  e^{\mathrm{i}\frac{\hat{H}_A+\hat{H}_B}
  {\hbar}(t_2-t_1)
  } \notag \\
  &=\hat{\rho}(t_1)
  \quad\mathrm{for}\quad
  t_1\leq t_2\leq0.
\end{align}
At time $t=0$,
we remove the constraint $\hat{H}_I=0$
and let the state of the total system
evolve under the total
Hamiltonian \eqref{eq:totH1}.
The interaction sets in
between the system and the bath,
which creates correlations.

As $\hat{H}_A$ and $\hat{H}_B$ are
purely quadratic,
the initial state \eqref{eq:ini2} is a Gaussian
state \cite{serafini2017quantum,doi:10.1142/S1230161214400010,RevModPhys.84.621,WANG20071,ferraro2005gaussian}
with vanishing first
moments: $\mathrm{Tr}\left[\hat{\bf{r}}\hat{\rho}(0)\right]
=\bf{0}$.
Moreover, as the total Hamiltonian is
purely quadratic, the total density operator
\begin{align}
  \hat{\rho}(t)=\hat{U}(t)\hat{\rho}(0)
  \hat{U}^{\dag}(t)
  \quad\mathrm{with}\quad
  \hat{U}(t)=\exp\left(
  -\mathrm{i}\frac{\hat{H}}{\hbar}t
  \right)
  \label{eq:unitary1}
\end{align}
is always a Gaussian state with vanishing first
moments: $\mathrm{Tr}\left[\hat{\bf{r}}\hat{\rho}(t)\right]
=\bf{0}$.
Therefore, $\hat{\rho}(t)$ is completely characterized
by the $2(N+1)\times 2(N+1)$ covariance matrix $\sigma(t)$
whose $(j,k)$-element is given by
\begin{align}
  \sigma_{j,k}(t)=
  \mathrm{Tr}\left[\hat{\rho}(t)\{\hat{r}_j,\hat{r}_k\}\right],
\end{align}
where the curly parentheses $\{\bullet,\bullet\}$ denote the anticommutator.
Note that the covariance matrix is a symmetric matrix.
Because of Eq.~\eqref{eq:unitary1},
the following relation holds
\cite[Sec.~5.1.2]{serafini2017quantum}:
\begin{align}
  \sigma(t)&=V(t)\sigma(0)V(t)^{\mathrm{T}}
  \quad\mathrm{with}\quad
  V(t)=\mathrm{e}^{\Omega Ht},
  \label{eq:unitary2}
\end{align}
where
\begin{align}
  \Omega&=\bigoplus_{j=1}^{N+1}\Omega_1
  =\left(\begin{array}{ccc}
  \Omega_1 & & \\
  & \ddots & \\
   & & \Omega_1
  \end{array}\right)
   , \quad
  \Omega_1=
  \left(\begin{array}{cc}0 & 1 \\
  -1 & 0 \\
  \end{array}\right),
  \label{eq:Omega}
\end{align}
and $H$ is the $2(N+1)$-dimensional symmetric matrix
introduced in Eq.~\eqref{eq:totH1}.

If the total system is in a Gaussian state,
its subsystems are also in Gaussian states.
Thus, each of the states of the system and the bath is
Gaussian and is completely characterized by
the covariance matrices $\sigma_A(t)$
and $\sigma_B(t)$,
respectively,
which are the submatrices
of $\sigma(t)$
\cite[Sec.~5.2]{serafini2017quantum}:
\begin{align}
  \sigma(t)=
  \begin{pmatrix}
    \sigma_A(t) & \sigma_{AB}(t) \\
    \sigma_{AB}(t)^{\mathrm{T}} & \sigma_B(t) \\
  \end{pmatrix},
  \label{eq:submatrices1}
\end{align}
where $\sigma_A(t)$ is a two-dimensional
symmetric matrix, $\sigma_B(t)$ is a $2N$-dimensional
symmetric matrix, and $\sigma_{AB}(t)$ is a
$2\times 2(N+1)$ matrix.
Each harmonic oscillator in the total system
is also in a Gaussian state which
is totally determined by the following
covariance matrix:
\begin{align}
  \sigma_j(t):=
  \begin{pmatrix}
    \sigma_{2j-1,2j-1}(t) & \sigma_{2j-1,2j}(t) \\
    \sigma_{2j-1,2j}(t) & \sigma_{2j,2j}(t) \\
  \end{pmatrix}
  \label{eq:sigmajt1}
\end{align}
for $j=1,\dots,N+1$.
The initial covariance matrix
for the state \eqref{eq:ini2} is
\cite[Sec.~3.3]{serafini2017quantum}
\begin{align}
  \begin{gathered}
    \sigma(0)=
    \begin{pmatrix}
      \sigma_A(0) & 0 \\
      0 & \sigma_B(0) \\
    \end{pmatrix},
    \\
    \sigma_A(0)=\sigma_1(0)=
    \coth\left(
    \frac{\hbar\omega_1}{2k_BT_A^0}
    \right)I_2,
    \\
    \sigma_B(0)=\bigoplus_{j=2}^{N+1}
    \sigma_j(0),\quad
    \sigma_j(0)=
    \coth\left(
    \frac{\hbar\omega_{j}}{2k_BT_B^0}
    \right)I_2,
  \end{gathered}
  \label{eq:inicov1}
\end{align}
where $T_A^0=1/(k_B\beta_A^0)$,
$T_B^0=1/(k_B\beta_B^0)$, and
$I_2$ is the two-dimensional
identity matrix.

\subsection{The GKSL master equation}
\label{GKSL}
If the couplings $\{g_j\}$ of
the harmonic oscillators are sufficiently weak,
the dynamics of the system is well approximated by the
GKSL master equation \cite{breuer2002theory,rivas2012open,Rivas_2010,doi:10.1063/1.522979,lindblad1976generators}:
\begin{align}
  \frac{\mathrm{d}}{\mathrm{d}t}\hat{\rho}_{A}(t)
  &=-\frac{\mathrm{i}}{\hbar}
  \left[
  \hat{H}_A,\hat{\rho}_{A}(t)
  \right]\notag\\
  &\quad+\Gamma(\bar{n}+1)
  \left(2\hat{a}_1\hat{\rho}_{A}(t)\hat{a}_1^{\dag}
  -\left\{\hat{a}_1^{\dag}\hat{a}_1,\hat{\rho}_{A}(t)\right\}\right)
  \notag \\
  &\quad\mbox{}+\Gamma\bar{n}
  \left(2\hat{a}_1^{\dag}\hat{\rho}_{A}(t)\hat{a}_1
  -\left\{\hat{a}_1\hat{a}_1^{\dag},\hat{\rho}_{A}(t)
  \right\}\right),
  \label{eq:GKSL1}
\end{align}
where
\begin{align}
  \bar{n}=\frac{1}{\mathrm{e}^{\beta_B^0\hbar\omega_1}-1}
\end{align}
is the mean excitation number of a
harmonic oscillator at thermal equilibrium
with frequency $\omega_1$ at inverse temperature $\beta_B^0$, and
\begin{align}
  \Gamma=\pi J(\omega_1)
  \label{eq:Gamma1}
\end{align}
is the relaxation rate of the system, with
\begin{align}
  J(\omega)=\sum_{j=2}^{N+1}g_j^2\ \delta(\omega-\omega_j)
\end{align}
being the spectral density of the bath.
Note that, when we calculate $\Gamma$ in
Eq.~\eqref{eq:Gamma1}, we need to specify the form of $J(\omega)$ in the continuous limit.
For example, if we consider an Ohmic bath
\cite{Rivas_2010,RevModPhys.89.015001}, the spectral density is written as 
$J(\omega)=\eta\omega\mathrm{e}^{-\omega/\omega_c}$,
where $\eta$ is the coupling strength between the system and the bath, and $\omega_c$ is the cutoff frequency.
Under this GKSL master equation,
the system is equilibrated with the bath
in the limit $t\to\infty$:
\begin{align}
    \hat{\rho}_{A}(\infty)
    =\hat{\rho}_{A}^{\mathrm{th}}=
    \frac{\mathrm{e}^{-\beta_B^0\hat{H}_A}}
    {\mathrm{Tr}[\mathrm{e}^{-\beta_B^0\hat{H}_A}]}
\end{align}

As we will show in the next section,
$[\hat{H}_A,\hat{\rho}_{A}(t)]=0$
always holds in our settings.
Then the GKSL master equation \eqref{eq:GKSL1}
becomes
\begin{align}
  \frac{\mathrm{d}}{\mathrm{d}t}
  \hat{\rho}_{A}(t)
  &=\Gamma(\bar{n}+1)
  \left(2\hat{a}_1\hat{\rho}_{A}(t)\hat{a}_1^{\dag}
  -\left\{\hat{a}_1^{\dag}\hat{a}_1,\hat{\rho}_{A}(t)\right\}\right)\notag\\
  &\quad+\Gamma\bar{n}
  \left(2\hat{a}_1^{\dag}\hat{\rho}_{A}(t)\hat{a}_1
  -\left\{\hat{a}_1\hat{a}_1^{\dag},\hat{\rho}_{A}(t)
  \right\}\right)
  \label{eq:GKSL2}
\end{align}
Under this GKSL master equation
and the initial covariance matrix
in Eq.~\eqref{eq:inicov1},
the covariance matrix
of the system at time $t$
is written as \cite[Sec.~4.1.1]{ferraro2005gaussian}
\begin{align}
  \sigma_A(t)
  &=\left[\coth\left(
  \frac{\hbar\omega_1}{2k_BT_A^0}
  \right)\mathrm{e}^{-2\Gamma t}
  \right.
  \notag \\
  &\quad\mbox{}+\left.\coth\left(
  \frac{\hbar\omega_1}{2k_BT_B^0}
  \right)\left(1-\mathrm{e}^{-2\Gamma t}
  \right)\right]I_2.
  \label{eq:sigmaAt1}
\end{align}

\section{Analytical results}
\label{sec:results1}
\subsection{Gibbs states}
\label{subsec:Gibbs}
We show that each harmonic oscillator is always
in a Gibbs state with a time-dependent temperature
under the unitary dynamics \eqref{eq:unitary1}
of the total system.
Note that there is a one-to-one
correspondence between the density operator
and the covariance matrix of each harmonic oscillator.
As the covariance matrix is easier to calculate
than the density matrix,
we first calculate the covariance matrix.
By substituting Eq.~\eqref{eq:inicov1}
into Eq.~\eqref{eq:unitary2}, we find
(see Appendix~\ref{app:Gibbs1})
\begin{align}
  \sigma_j(t)=\sigma_{2j-1,2j-1}(t)I_2
  \quad\mathrm{for}\quad
  j=1,\dots,N+1.
  \label{eq:sigmajt2}
\end{align}
According to the calculation
in Appendix~\ref{app:Gibbs1},
the density operator is expressed
with the covariance matrix \eqref{eq:sigmajt2}
in the following form:
\begin{align}
  \hat{\rho}_j(t)&=
  \frac{\mathrm{e}^{-\beta_j(t)\hat{H}_j}}
  {Z_j(t)},
  \label{eq:rhojt2}\\
  Z_j(t)&=\mathrm{Tr}\left[
  \mathrm{e}^{-\beta_j(t)\hat{H}_j}
  \right]
  =\frac{1}{2}\sqrt{
  \sigma_{2j-1,2j-1}(t)^2-1
  },
  \label{eq:Zjt2}
  \\
  \beta_j(t)&=\frac{1}{k_BT_j(t)}
  =\frac{2}{\hbar\omega_j}
  \coth^{-1}\left[
  \sigma_{2j-1,2j-1}(t)
  \right]\notag\\
  &=\frac{1}{\hbar\omega_j}
  \ln\left(
  \frac{\sigma_{2j-1,2j-1}(t)+1}
  {\sigma_{2j-1,2j-1}(t)-1}
  \right)
  \notag \\
  &=\frac{1}{\hbar\omega_j}
  \ln\left(
  \frac{2E_j(t)+\hbar\omega_j}
  {2E_j(t)-\hbar\omega_j}
  \right),
  \label{eq:Tjt1}
\end{align}
where $E_j(t)$ is the mean energy of
the $j$th harmonic oscillator:
\begin{align}
  E_j(t)=\mathrm{Tr}
  \left[
  \hat{H}_j\hat{\rho}_j(t)
  \right]
  =\frac{\hbar\omega_j}{2}
  \sigma_{2j-1,2j-1}(t).
  \label{eq:Ejt1}
\end{align}
We find that each harmonic oscillator
is always in a Gibbs (thermal equilibrium) state
with a time-dependent temperature $T_j(t)$.
In this meaning, the dynamics is quasistatic
for every harmonic oscillator.

As the system is always in a Gibbs state,
the relation
$[\hat{H}_A,\hat{\rho}_{A}(t)]=[
\hat{H}_A,\mathrm{e}^{-\beta_A(t)\hat{H}_A}/Z_A(t)]=0$ holds all the time.
Therefore the GKSL master equation
\eqref{eq:GKSL1} transforms into
Eq.~\eqref{eq:GKSL2}.
Using Eq.~\eqref{eq:sigmaAt1}
for the time-dependent temperature $T_j(t)$
in Eq.~\eqref{eq:Tjt1},
we find that the system
under the GKSL master equation
is equilibrated with the bath
in the limit $t\to\infty$:
\begin{align}
  T_A(\infty)=T_1(\infty)=T_B^0;
\end{align}
we will plot this
in Fig.~\ref{fig:temperature1} below.

\subsection{Thermodynamic entropy}
\label{subsec:Def1}
We define
the time-dependent
free energy and the time-dependent
thermodynamic entropy
of the $j$th harmonic oscillator
simply following 
the analog of equilibrium
statistical mechanics and
thermodynamics:
\begin{align}
  F_j(t)&:=-k_BT_j(t)\ln Z_j(t), \\
  S_j^{\mathrm{th}}(t)
  &:=\frac{E_j(t)-F_j(t)}{T_j(t)}.
  \label{eq:defSjth1}
\end{align}
In fact, the von Neumann entropy
of the $j$th harmonic oscillator
coincides with its thermodynamic entropy
because it is in a Gibbs state
\cite[Sec.~21.1]{oono2017perspectives}:
\begin{align}
  S_j^{\mathrm{vN}}(t)
  &:=-k_B\mathrm{Tr}
  \left[
  \hat{\rho}_j(t)\ln\hat{\rho}_j(t)\right]\notag\\
  &=-k_B\mathrm{Tr}
  \left[\hat{\rho}_j(t)
  \ln\left(\frac{\mathrm{e}^{-\beta_j(t)\hat{H}_j}}
  {Z_j(t)}
  \right)
  \right]
  \notag \\
  &=\frac{1}{T_j(t)}
  \mathrm{Tr}
  \left[
  \hat{\rho}_j(t)
  \hat{H}_j
  \right]
  +k_B\ln Z_j(t)\notag\\
  &=
  \frac{E_j(t)-F_j(t)}{T_j(t)}
  =S_j^{\mathrm{th}}(t).
  \label{eq:vonent1}
\end{align}

We can rewrite $S_j^{\mathrm{th}}(t)$
in Eq.~\eqref{eq:defSjth1} as a strictly
monotonically increasing function of $E_j(t)$:
\begin{align}
  \frac{S_j^{\mathrm{th}}(t)}{k_B}
  &=\frac{2E_j(t)+\hbar\omega_j}{2\hbar\omega_j}
  \ln\left(
  \frac{2E_j(t)+\hbar\omega_j}{2\hbar\omega_j}
  \right)\notag\\
  &\qquad-
  \frac{2E_j(t)-\hbar\omega_j}{2\hbar\omega_j}
  \ln\left(
  \frac{2E_j(t)-\hbar\omega_j}{2\hbar\omega_j}
  \right).
  \label{eq:thent1}
\end{align}
This is followed by
\begin{align}
  \frac{\partial S_j^{\mathrm{th}}(t)}
  {\partial E_j(t)}
  =\frac{k_B}{\hbar\omega_j}
  \ln\left(
  \frac{2E_j(t)+\hbar\omega_j}{2E_j(t)-\hbar\omega_j}
  \right)
  =\frac{1}{T_j(t)}
\end{align}
and
\begin{align}
  \frac{\mathrm{d}}{\mathrm{d}t}
  S_j^{\mathrm{th}}(t)
  =\frac{1}{T_j(t)}
  \frac{\mathrm{d}}{\mathrm{d}t}
  E_j(t).
  \label{eq:quasi_jth}
\end{align}
We regard $\mathrm{d}E_j(t)/\mathrm{d}t$
as the heat flux into the $j$th harmonic oscillator
because its Hamiltonian $\hat{H}_j$ is time independent
\cite[Sec.~2.1]{doi:10.1080/00107514.2016.1201896}.
Then, Eq.~\eqref{eq:quasi_jth} is a manifestation of the quasi-static process;
see Eq.~\eqref{eq:dSAdt1}.
We define the thermodynamic entropy
flux into the $j$th harmonic oscillator as
\begin{align}
    \frac{\mathrm{d}^{\mathrm{ext}}}{\mathrm{d}t}
    S^{\mathrm{th}}_j(t)
    &=\frac{1}{T_j(t)}\frac{\mathrm{d}}{\mathrm{d}t}
    E_j(t),
    \label{eq:efluxj1}
\end{align}
just as Eq.~\eqref{eq:efluxA1}.
Then we find that the internal thermodynamic entropy
production rate of the $j$th harmonic oscillator
is zero:
\begin{align}
    \frac{\mathrm{d}^{\mathrm{int}}}{\mathrm{d}t}
    S^{\mathrm{th}}_j(t)
    =\frac{\mathrm{d}}{\mathrm{d}t}
    S^{\mathrm{th}}_j(t)
    -\frac{\mathrm{d}^{\mathrm{ext}}}{\mathrm{d}t}
    S^{\mathrm{th}}_j(t)
    =0,
\end{align}
which is also a manifestation of the quasistatic process.

In order to define the nonequilibrium thermodynamic
entropy of the total system,
we impose the additivity
of the thermodynamic entropy,
which is satisfied in equilibrium
thermodynamics of macroscopic systems
(see Secs.~11.5 and 13.11 in
Ref.~\cite{oono2017perspectives}).
We thereby arrive at
\begin{align}
  S_{\mathrm{tot}}^{\mathrm{th}}(t)
  &:=\sum_{j=1}^{N+1}
  S_j^{\mathrm{th}}(t)
  \notag \\
  &=k_B\sum_{j=1}^{N+1}
  \left[
  \frac{2E_j(t)+\hbar\omega_j}{2\hbar\omega_j}
  \ln\left(
  \frac{2E_j(t)+\hbar\omega_j}{2\hbar\omega_j}
  \right)\right.\notag\\
  &\left.\qquad-
  \frac{2E_j(t)-\hbar\omega_j}{2\hbar\omega_j}
  \ln\left(
  \frac{2E_j(t)-\hbar\omega_j}{2\hbar\omega_j}
  \right)
  \right].
  \label{eq:thent3}
\end{align}

We analytically confirm
that our thermodynamic entropy \eqref{eq:thent3} satisfies
the third law of thermodynamics
\cite[Sec.~23.7]{oono2017perspectives} as follows.
The temperature $T_j(t)$ in Eq.~\eqref{eq:Tjt1}
and the thermodynamic entropy $S_j^{\mathrm{th}}(t)$
in Eq.~\eqref{eq:thent1}
become zero for the vacuum state:
\begin{align}
  T_j(t)\to+0, \quad
  S_j^{\mathrm{th}}(t)\to+0
  \quad\mathrm{as}\quad
  E_j(t)\to \frac{\hbar\omega_j}{2}+0.
  \label{eq:lim1}
\end{align}
As $T_j(t)$ and $S_j^{\mathrm{th}}(t)$
are both strictly monotonically increasing functions
of $E_j(t)$,
the thermodynamic entropy $S_j^{\mathrm{th}}(t)$
becomes zero if and only if $T_j(t)$ becomes zero:
\begin{align}
  S_j^{\mathrm{th}}(t)\to+0
  \quad\mathrm{as}\quad
  T_j(t)\to+0.
  \label{eq:lim2}
\end{align}
This and Eq.~\eqref{eq:thent3}
lead to the third law of thermodynamics:
\begin{align}
  S_{\mathrm{tot}}^{\mathrm{th}}(t)\to+0
  \quad\mathrm{as}\quad
  T_j(t)\to+0 \quad^{\forall}j,
  \label{eq:third}
\end{align}
which supports the validity
of our definition of the total thermodynamic
entropy in Eq.~\eqref{eq:thent3}.

\subsection{Total thermodynamic entropy production and its rate}
We define the total thermodynamic
entropy production as
\begin{align}
  \Delta S_{\mathrm{tot}}^{\mathrm{th}}(t)
  &:=S_{\mathrm{tot}}^{\mathrm{th}}(t)
  -S_{\mathrm{tot}}^{\mathrm{th}}(0)\notag\\
  &=S_A^{\mathrm{th}}(t)-S_A^{\mathrm{th}}(0)
  +\sum_{j=2}^{N+1}\left[
  S_j^{\mathrm{th}}(t)-S_j^{\mathrm{th}}(0)
  \right]
  \label{eq:TEP1}
\end{align}
and its rate as
\begin{align}
  \Pi_{\mathrm{tot}}^{\mathrm{th}}(t)
  &:=\frac{\mathrm{d}}{\mathrm{d}t}
  S_{\mathrm{tot}}^{\mathrm{th}}(t)
  =\sum_{j=1}^{N+1}
  \frac{1}{T_j(t)}
  \frac{\mathrm{d}}{\mathrm{d}t}
  E_j(t)\notag\\
  &=\frac{1}{T_A(t)}
  \frac{\mathrm{d}}{\mathrm{d}t}
  E_A(t)
  +\sum_{j=2}^{N+1}
  \frac{1}{T_j(t)}
  \frac{\mathrm{d}}{\mathrm{d}t}
  E_j(t).
  \label{eq:TEPR1}
\end{align}
Let us transform this into the form which
we can easily calculate in terms
of the covariance matrix.
Using Eqs.~\eqref{eq:dE1tdt1} and
\eqref{eq:dEjtdt1}
in Appendix~\ref{app:dEdt}, we obtain
\begin{align}
  \Pi_{\mathrm{tot}}^{\mathrm{th}}(t)
  &=\frac{\hbar\omega_1}{T_A(t)}
  \sum_{j=2}^{N+1}g_j
  \sigma_{1,2j}(t)
  -\sum_{j=2}^{N+1}
  \frac{\hbar\omega_j}{T_j(t)}
  g_j\sigma_{1,2j}(t)
  \notag \\
  &=k_B\sum_{j=2}^{N+1}
  g_j\sigma_{1,2j}(t)
  \left[
  \ln\left(
  \frac{\sigma_{1,1}(t)+1}
  {\sigma_{1,1}(t)-1}
  \right) \right.\notag\\
  &\left. \qquad-\ln\left(
  \frac{\sigma_{2j-1,2j-1}(t)+1}
  {\sigma_{2j-1,2j-1}(t)-1}
  \right)
  \right].
  \label{eq:TEPR2}
\end{align}
This total thermodynamic entropy production rate
can be negative as we will see later.

\subsection{The difference between
our total thermodynamic
entropy production rate and the conventional one}
\label{subsec:conventional}
Let us consider the weak-coupling regime
so that the dynamics of the system is well
approximated by the GKSL master equation in
Eq.~\eqref{eq:GKSL2}.
In our settings, the von Neumann entropy
of the system coincides with its thermodynamic
entropy as in Eq.~\eqref{eq:vonent1},
and hence the conventional
entropy production rate $\Pi^{\mathrm{vN}}(t)$
in Eq.~\eqref{eq:vonEPR2} has the following form:
\begin{align}
  \Pi^{\mathrm{vN}}(t)&=\frac{1}{T_A(t)}
  \frac{\mathrm{d}}{\mathrm{d}t}
  E_A(t)
  +\frac{1}{T_B^0}
  \frac{\mathrm{d}}{\mathrm{d}t}E_B(t)
  \notag \\
  &=\frac{1}{T_A(t)}
  \frac{\mathrm{d}}{\mathrm{d}t}
  E_A(t)
  +\sum_{j=2}^{N+1}\frac{1}{T_B^0}
  \frac{\mathrm{d}}{\mathrm{d}t}E_j(t).
  \label{eq:EPRpre1}
\end{align}
Let us transform Eq.~\eqref{eq:EPRpre1} into the form which
we can easily calculate.
As we consider the weak-coupling regime,
we neglect the interaction energy:
$\mathrm{d}E_B(t)/\mathrm{d}t=-\mathrm{d}E_A(t)/\mathrm{d}t$.
From the first line in Eq.~\eqref{eq:EPRpre1},
we obtain
\begin{align}
  \Pi^{\mathrm{vN}}(t)
  &=\left(
  \frac{1}{T_A(t)}-\frac{1}{T_B^0}
  \right)
  \frac{\mathrm{d}}{\mathrm{d}t}
  E_A(t)
  \notag \\
  &=\hbar\omega_1\Gamma
  \left(
  \frac{1}{T_B^0}-\frac{1}{T_A(t)}
  \right)
  \notag\\
  &\quad\times
  \left[
  \frac{2E_A(t)}{\hbar\omega_1}
  -\coth\left(
  \frac{\hbar\omega_1}{2k_BT_B^0}
  \right)
  \right]
  \notag \\
  &=\hbar\omega_1\Gamma
  \left(
  \frac{1}{T_B^0}-\frac{1}{T_A(t)}
  \right)
  \notag\\
  &\quad\times\left[
  \coth\left(
  \frac{\hbar\omega_1}{2k_BT_A(t)}
  \right)
  -\coth\left(
  \frac{\hbar\omega_1}{2k_BT_B^0}
  \right)
  \right],
  \label{eq:EPRpre2}
\end{align}
where the second line follows from
Eqs.~\eqref{eq:Ejt1} and \eqref{eq:sigmaAt1},
and the last line follows from
Eq.~\eqref{eq:Tjt1}.

The difference between our total thermodynamic
entropy production rate $\Pi_{\mathrm{tot}}^{\mathrm{th}}(t)$
in Eq.~\eqref{eq:TEPR1} and the conventional
entropy production rate $\Pi^{\mathrm{vN}}(t)$ in Eq.~\eqref{eq:EPRpre1}
arises from the gaps
between $\{T_j(t)\}$ and $T_B^0$:
\begin{align}
  \Pi^{\mathrm{vN}}(t)-\Pi_{\mathrm{tot}}^{\mathrm{th}}(t)
  =\sum_{j=2}^{N+1}\left(\frac{1}{T_B^0}
  -\frac{1}{T_j(t)}\right)
  \frac{\mathrm{d}}{\mathrm{d}t}E_j(t).
  \label{eq:difofepr1}
\end{align}

\section{Numerical results}
\label{sec:results2}
\subsection{Parameters}
\label{subsec:parameters}
For a numerical example,
we use an Ohmic
bath \cite{Rivas_2010,RevModPhys.89.015001},
whose spectral density is
\begin{align}
  J(\omega)
  =\sum_{j=2}^{N+1}g_j^2\delta(\omega-\omega_j)
  =\eta\omega
  \mathrm{e}^{-\omega/\omega_c},
  \label{eq:Jomega2}
\end{align}
where $\eta$ is the coupling strength
between the system and the bath,
and $\omega_c$ is the cutoff frequency.
For numerical demonstration,
we fix the parameters as follows
\cite[Appendix A]{Rivas_2010}:
\begin{gather}
  \omega_1=4\,\mathrm{MHz},\,
  \omega_c=3\,\mathrm{MHz},\,
  \omega_{\mathrm{min}}=0.026\,\mathrm{MHz},\notag\\
  \omega_{\mathrm{max}}=20\,\mathrm{MHz},\,
  \Delta\omega=\frac{\omega_{\mathrm{max}}
  -\omega_{\mathrm{min}}}{N-1},
  \notag \\
  \omega_j=
  \omega_{\mathrm{min}}
  +(j-2)\Delta\omega
  \quad\mathrm{for}\quad
  j=2,\dots,N+1,
  \notag \\
  \eta = 10^{-3},\,
  T_A^0=10\,\mu\mathrm{K},\,
  T_B^0=50\,\mu\mathrm{K}.
  \label{eq:paraOhmic1}
\end{gather}
We set the coupling constant $g_j$ by
integrating Eq.~\eqref{eq:Jomega2} over $\omega$ as in
\begin{align}
  \sum_{j=2}^{N+1}g_j^2
  =\int_{\omega_{\mathrm{min}}-\epsilon}
  ^{\omega_{\mathrm{max}}+\epsilon}
  \mathrm{d}\omega\,\eta\omega
  \mathrm{e}^{-\omega/\omega_c}
  \simeq
  \sum_{j=2}^{N+1}
  \eta\Delta\omega
  \omega_j
  \mathrm{e}^{-\omega_j/\omega_c},
\end{align}
which gives
\begin{align}
  g_j=\sqrt{\eta\Delta\omega
  \omega_j
  \mathrm{e}^{-\omega_j/\omega_c}}.
\end{align}

\begin{figure}
  \centering
  \includegraphics[width=0.45\textwidth]{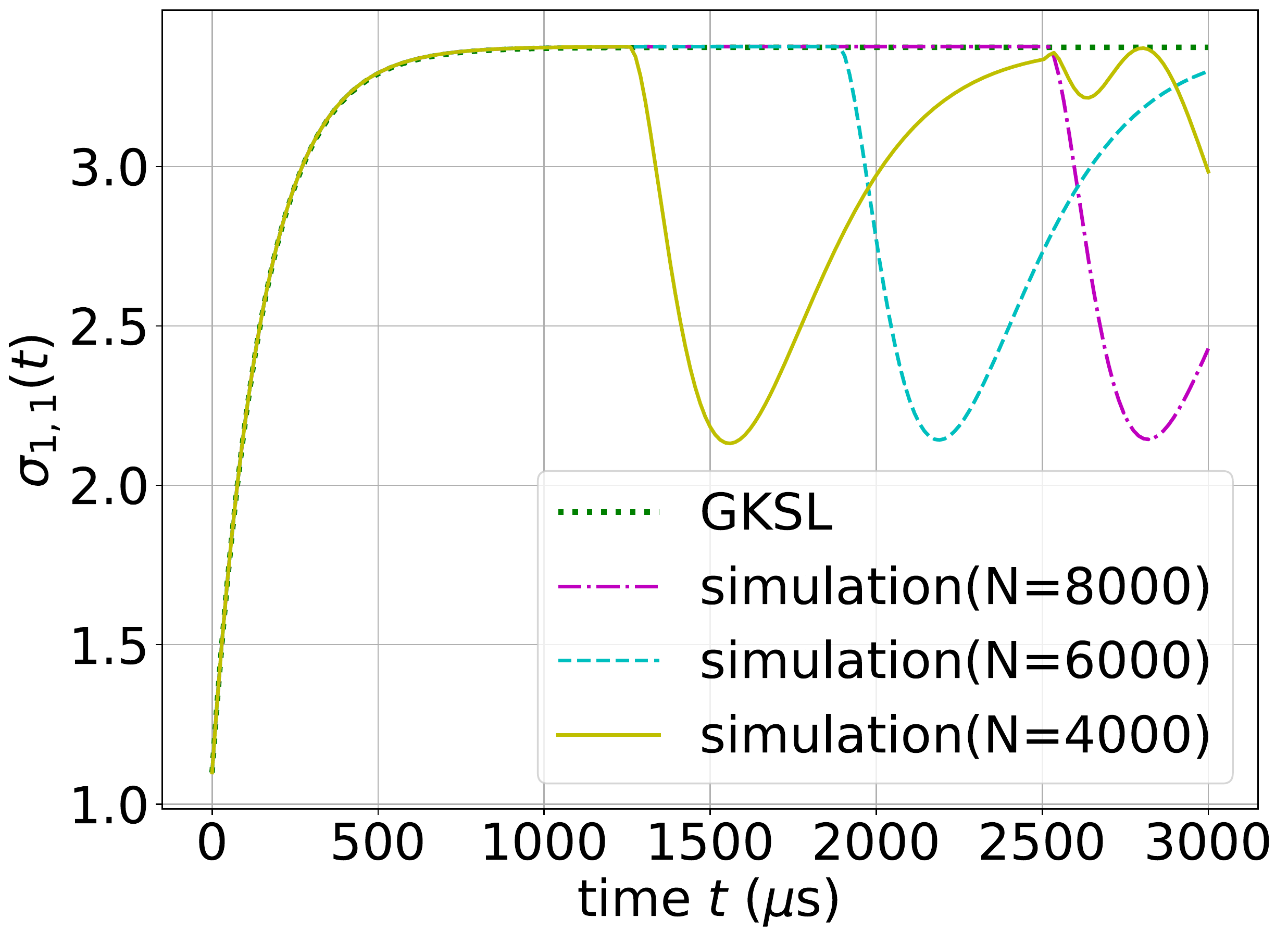}
  \caption{The time evolution of $\sigma_{1,1}(t)$.
  We set
  $\omega_1=4\,\mathrm{MHz}$,
  $\omega_c=3\,\mathrm{MHz}$,
  $\omega_{\mathrm{min}}=0.026\,\mathrm{MHz}$,
  $\omega_{\mathrm{max}}=20\,\mathrm{MHz}$,
  $\eta = 10^{-3}$,
  $T_A^0=10\,\mu\mathrm{K}$, and
  $T_B^0=50\,\mu\mathrm{K}$.
  The green dotted line is obtained from
  the solution of the GKSL master
  equation \eqref{eq:sigmaAt1}.
  The other lines are obtained from
  the unitary dynamics of the total
  system \eqref{eq:unitary2}.}
  \label{fig:elements1}
\end{figure}

\begin{figure}
  \centering
  \includegraphics[width=0.45\textwidth]{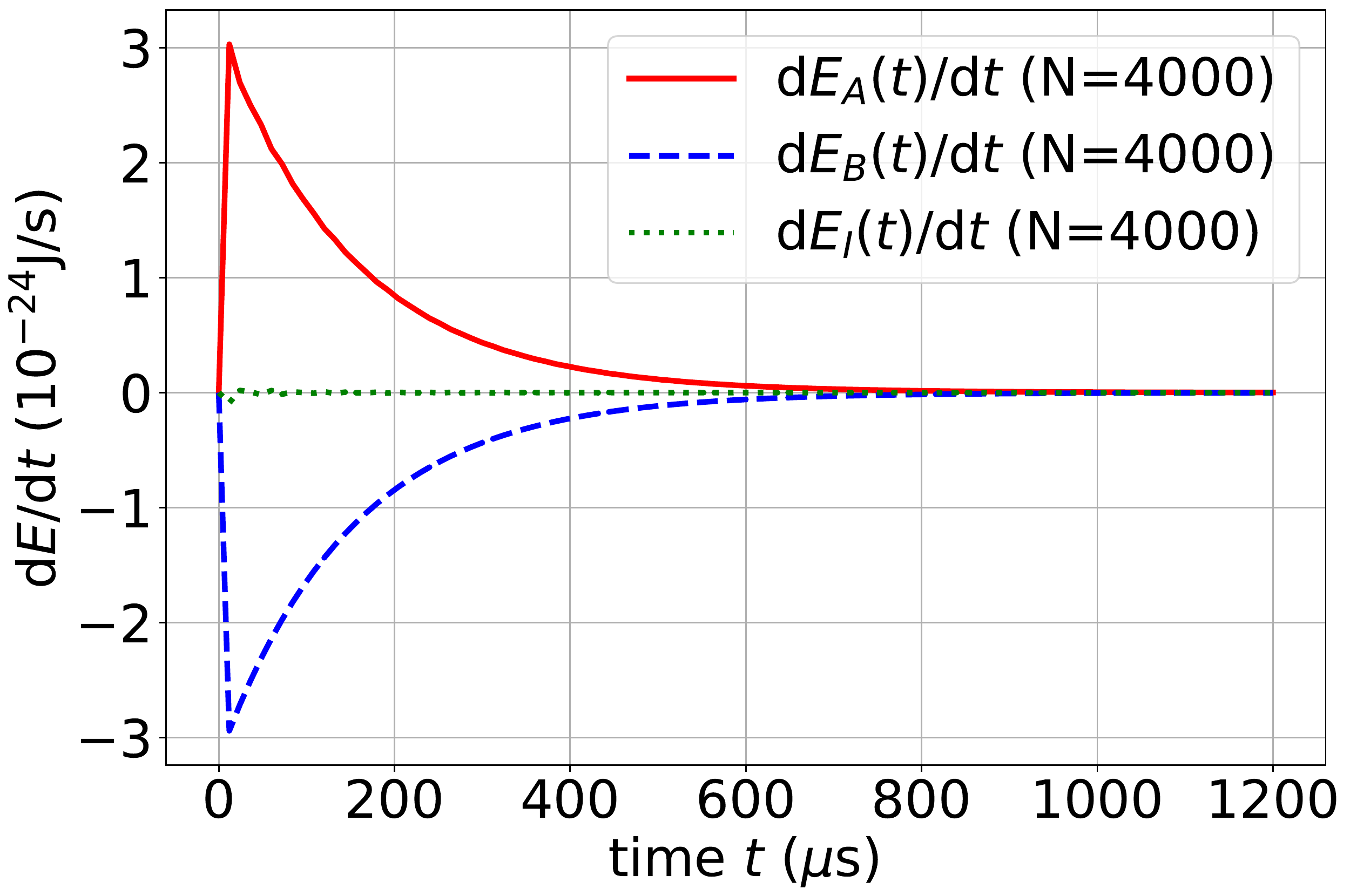}
  \caption{The time evolution of $\mathrm{d}E_A(t)/\mathrm{d}t$
  in Eq.~\eqref{eq:dE1tdt1}, $\mathrm{d}E_B(t)/\mathrm{d}t$
  in Eq.~\eqref{eq:dEBtdt1}, and $\mathrm{d}E_I(t)/\mathrm{d}t$
  in Eq.~\eqref{eq:dEItdt1} when $N=4000$
  under the unitary dynamics of the total
  system \eqref{eq:unitary2}.
  All the parameters except $N$ are the same as
  those in Fig.~\ref{fig:elements1}.}
  \label{fig:EABIt1}
\end{figure}

Let us check whether the dynamics of the system
obeys the GKSL master equation when $N=4000$,
$6000$, and $8000$.
Note that the quantum state of the system is totally
determined only by $\sigma_{1,1}(t)$.
Thus, in Fig.~\ref{fig:elements1} we compare $\sigma_{1,1}(t)$
which we calculate from the unitary dynamics of the total
system \eqref{eq:unitary2} and that
we calculate from the GKSL master
equation \eqref{eq:sigmaAt1}.
We find that the two curves coincide
with each other for $t\lesssim2\pi/\Delta\omega$, and hence
we conclude that the
dynamics of the system
is well approximated by the
GKSL master equation in that time range.
However, the dynamics of the system no longer obeys
the GKSL master equation
for $t\gtrsim2\pi/\Delta\omega$
because at $t=t_1:=2\pi/\Delta\omega$,
we have
$\mathrm{e}^{\mathrm{i}\omega_jt_1}=
\mathrm{e}^{2\pi\mathrm{i}\omega_{\mathrm{min}}/\Delta\omega}$
for $j=2,\dots,N+1$, and hence all harmonic oscillators
in the bath have almost the same phase
and recurrencelike behavior happens;
see Fig.~\ref{fig:elements1}.
Hence we restrict ourselves to
$t_{\mathrm{max}}<t_1=2\pi/\Delta\omega$
in the following calculations.
Note that $t_1$ is almost proportional to $N$
for large $N$ because
$\Delta\omega=(\omega_{\mathrm{max}}-\omega_{\mathrm{min}})
/(N-1)$; we thus need not worry about the
recurrencelike behavior for sufficiently
large $N$.
We also remark that the interaction energy
$E_I(t):=\mathrm{Tr}[\hat{\rho}(t)\hat{H}_I]$ is negligibly
small under the parameters in Eq.~\eqref{eq:paraOhmic1}
for large $N$;
see Fig.~\ref{fig:EABIt1}.
This justifies the transformation from the first line of
Eq.~\eqref{eq:EPRpre1} to that of Eq.~\eqref{eq:EPRpre2}.

\subsection{Negative total thermodynamic entropy production rate}
\label{subsec:tworates}
\begin{figure}
  \centering
  \includegraphics[width=0.45\textwidth]{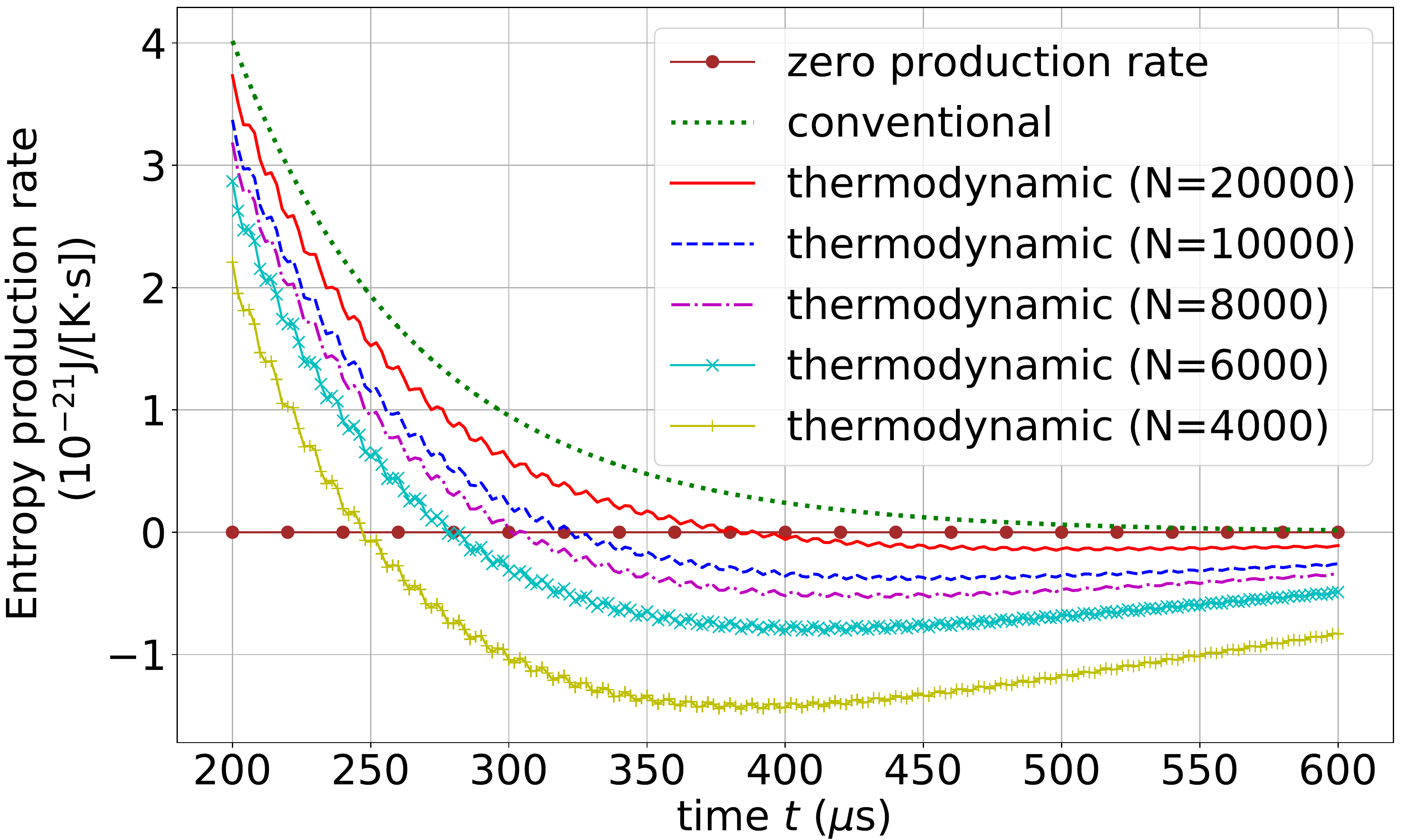}
  \caption{The total thermodynamic entropy
  production rate $\Pi_{\mathrm{tot}}^{\mathrm{th}}(t)$
  in Eq.~\eqref{eq:TEPR2}
  and the conventional entropy
  production rate ${\Pi}^{\mathrm{vN}}(t)$ in Eq.~\eqref{eq:EPRpre2}.
  All the parameters except $N$ are the same as
  those in Fig.~\ref{fig:elements1}.}
  \label{fig:EPR1}
\end{figure}
We compare in Fig.~\ref{fig:EPR1} our total thermodynamic
entropy production rate $\Pi_{\mathrm{tot}}^{\mathrm{th}}(t)$
in Eq.~\eqref{eq:TEPR2}
with the conventional
entropy production rate ${\Pi}^{\mathrm{vN}}(t)$ in Eq.~\eqref{eq:EPRpre2}.
We find that our total thermodynamic entropy production
rate $\Pi_{\mathrm{tot}}^{\mathrm{th}}(t)$
is negative in a certain time range, in contrast to
the conventional entropy production rate ${\Pi}^{\mathrm{vN}}(t)$,
which is always non-negative.
As we said in Sec.~\ref{subsec:conventional},
$\Pi_{\mathrm{tot}}^{\mathrm{th}}(t)$
differs from ${\Pi}^{\mathrm{vN}}(t)$
because some of $\{T_j(t)\}$ differ from $T_B^0$;
see Eq.~\eqref{eq:difofepr1}.
Let us see the behaviors of $\{T_j(t)\}$
below.

\begin{figure}
  \centering
  \includegraphics[width=0.45\textwidth]{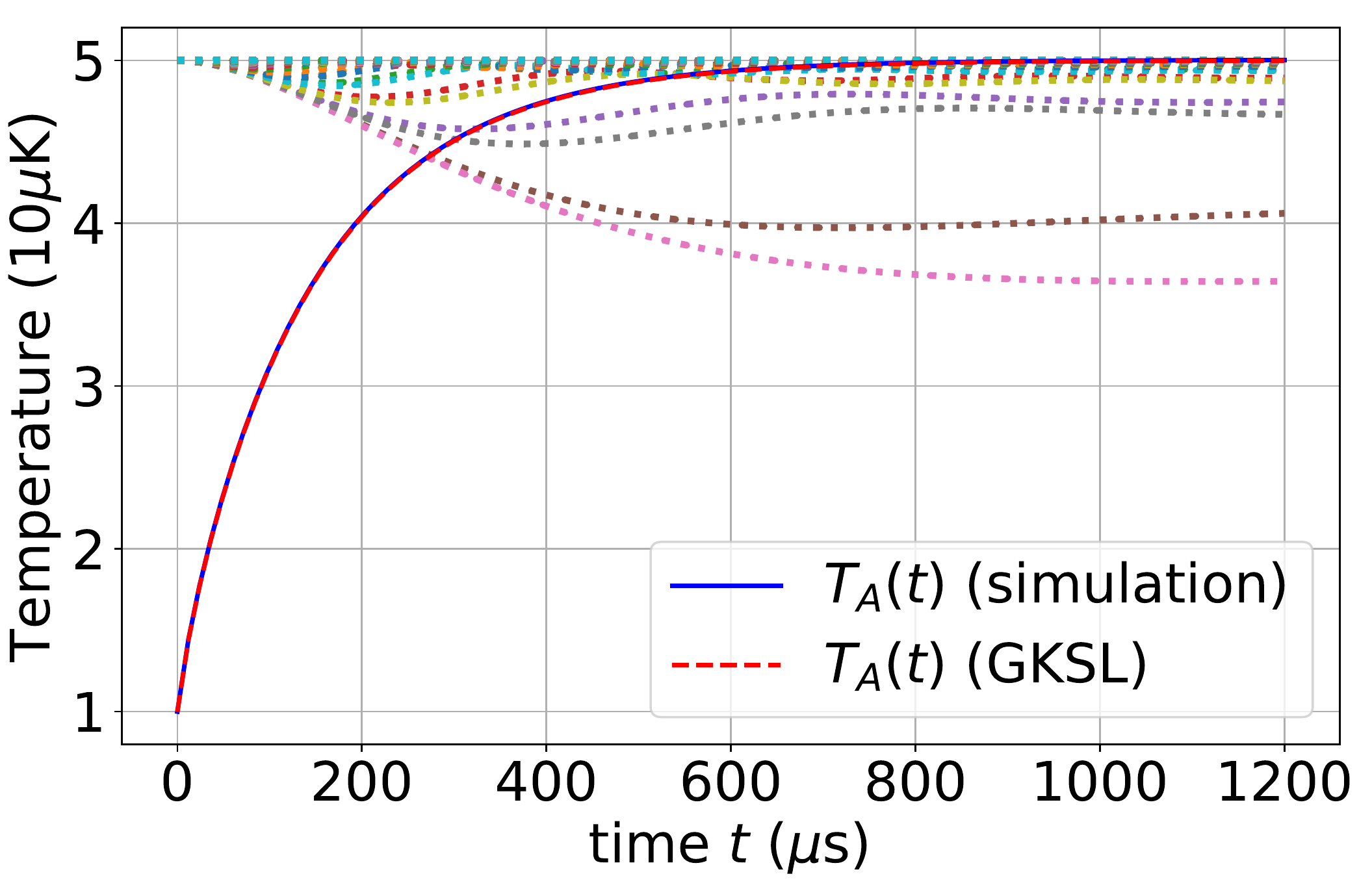}
  \caption{Time-dependent temperature $T_j(t)$
  in Eq.~\eqref{eq:Tjt1}
  of each harmonic oscillator.
  We set $N=4000$.
  All the parameters except $N$
  are the same
  as those in Fig.~\ref{fig:elements1}.
  The blue solid line is the time-dependent temperature
  of the system obtained from
  the unitary dynamics of the total
  system \eqref{eq:unitary2}.
  The red dashed line, which is almost identical to the blue solid line,
  is the time-dependent temperature
  of the system obtained from the
  solution of the GKSL master
  equation \eqref{eq:sigmaAt1}.
  The dotted lines are the time-dependent temperatures
  of all the harmonic oscillators in the bath obtained from the unitary dynamics of the total system.}
  \label{fig:temperature1}
\end{figure}
\begin{figure*}
 \centering
  \includegraphics[width=\textwidth]{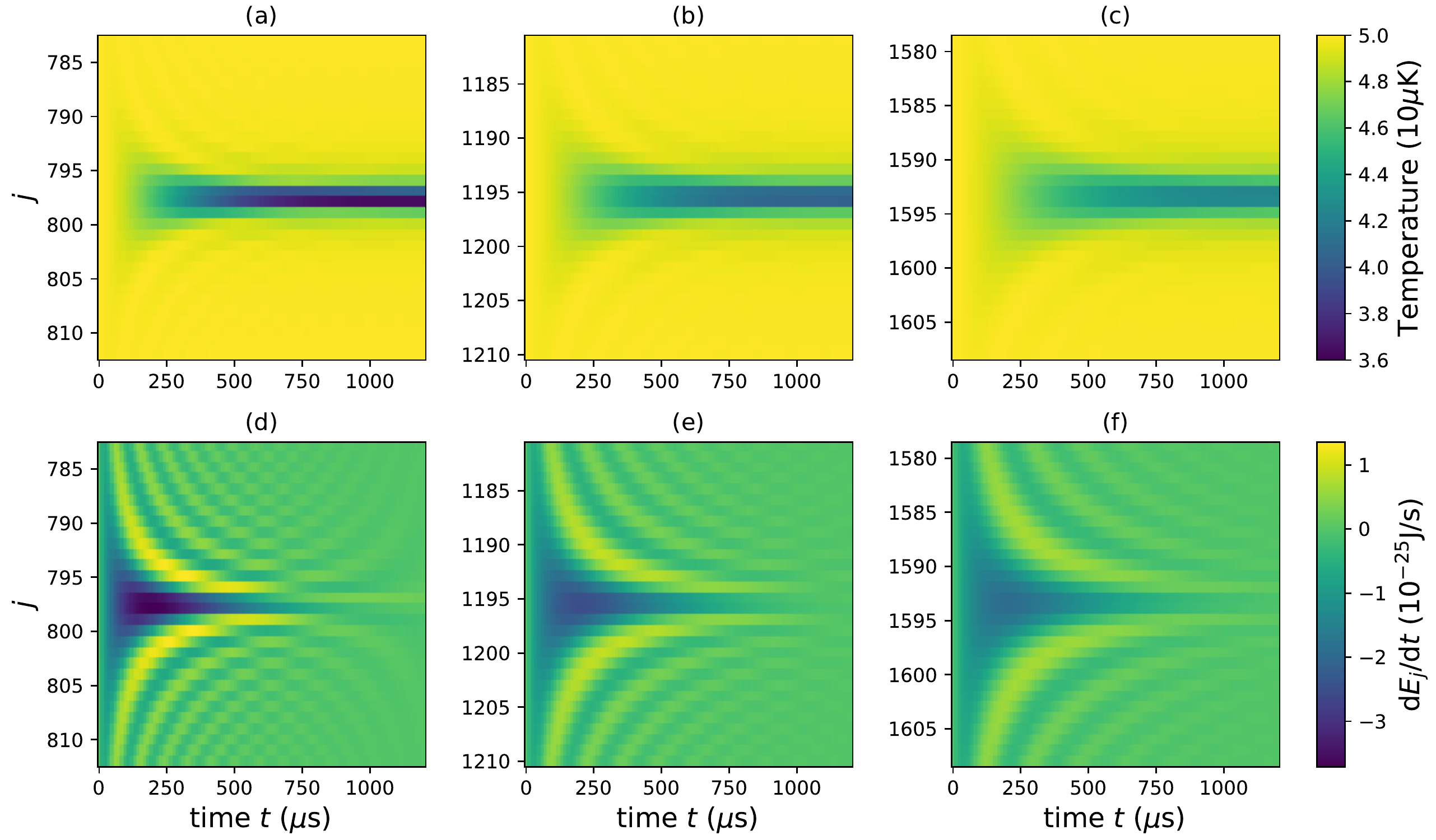}
  \caption{The time-dependent
  temperatures $\{T_j(t)\}$
  (upper panels) and
  the time derivatives of the mean energies
  $\{\mathrm{d}E_j(t)/\mathrm{d}t\}$
  (lower panels)
  of the harmonic oscillators
  in the bath which have almost the same
  frequencies as that of the system.
  The color expresses the value
  of $T_j(t)$ ($\mathrm{d}E_j(t)/\mathrm{d}t$)
  in the upper (lower) panels.
  The vertical axis corresponds
  to the number $j$ of each harmonic oscillator.
  The horizontal axis corresponds to time.
  We set $N=4000$ in (a) and (d), $N=6000$
  in (b) and (e), and $N=8000$ in (c) and (f).
  The other parameters are the same as
  those in Fig.~\ref{fig:elements1}.}
  \label{fig:temperature2}
\end{figure*}

We find in Fig.~\ref{fig:temperature1}
that the temperature of the system
$T_A(t)$ relaxes to the initial
temperature of the bath $T_B^0$,
while some of the
temperatures $\{T_j(t)\}$ of the harmonic oscillators
in the bath decrease.
The harmonic oscillators which show temperature 
decreasing have almost the same
frequency as the system
(Fig.~\ref{fig:temperature2}).
This can be explained as follows.
The mean energy of the system $E_A(t)$
is a strictly monotonically increasing function
of the temperature of the system $T_A(t)$,
and hence $E_A(t)$ increases as $T_A(t)$
relaxes to $T_B^0$, which is higher than
the initial temperature of the system $T_A^0$.
In order for $E_A(t)$ to increase,
the system must receive particles
with energy $\hbar\omega_1$.
Note that the total particle number operator
$\sum_{j=1}^{N+1}\hat{a}^{\dag}_j\hat{a}_j$
commutes with the total Hamiltonian
\eqref{eq:totH1}, so that the total
particle number is conserved.
Thus, in order for the system to
receive a particle with energy $\hbar\omega_1$,
the bath must provide the particle,
and only the harmonic oscillators
whose frequencies are almost the same as the
system can do so.
When the harmonic oscillators provide
the particle, their mean energies
$\{E_j(t)\}$ decrease.
Hence, 
the time-dependent
temperature $T_j(t)$, which is
a strictly monotonically increasing function
of $E_j(t)$,
also decreases.

We see from 
Fig.~\ref{fig:temperature2} that as
$|\omega_j-\omega_1|$ becomes smaller,
$[T_B^0-T_j(t)]$
and $|\mathrm{d}E_j(t)/\mathrm{d}t|$ become larger,
and so does $[1/T_B^0-1/T_j(t)]\mathrm{d}E_j(t)/\mathrm{d}t$.
As $N$ becomes larger, more
harmonic oscillators in the bath take part
in the energy exchange with the system,
and hence $[T_B^0-T_j(t)]$
and $|\mathrm{d}E_j(t)/\mathrm{d}t|$
for each harmonic oscillator become smaller;
see Fig.~\ref{fig:temperature2}.
In addition, $\sum_{j=2}^{N+1}\mathrm{d}E_j(t)/\mathrm{d}t
=\mathrm{d}E_B(t)/\mathrm{d}t=-\mathrm{d}E_A(t)/\mathrm{d}t$
does not depend on $N$ as long as the dynamics
of the system obeys the GKSL master equation.
Therefore as $N$ becomes larger,
${\Pi}^{\mathrm{vN}}(t)-\Pi_{\mathrm{tot}}^{\mathrm{th}}(t)$
in Eq.~\eqref{eq:difofepr1}
becomes smaller as in Fig.~\ref{fig:EPR1}.

\subsection{The second law of thermodynamics}
\label{sec:second}
\begin{figure}
  \centering
    \includegraphics[width=0.45\textwidth]{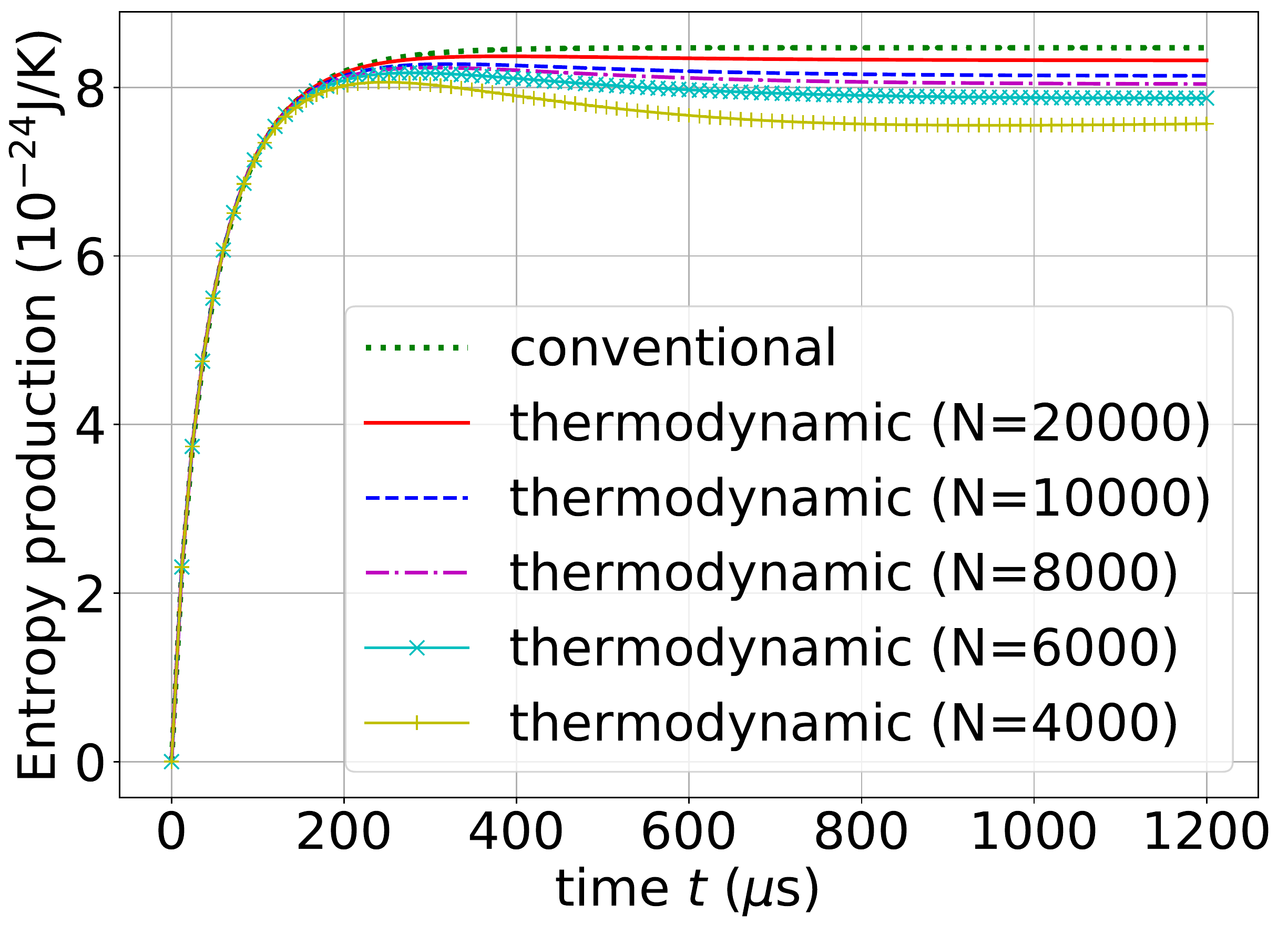}
    \caption{The thermodynamic entropy
    production $\Delta S_{\mathrm{tot}}^{\mathrm{th}}(t)$
    in Eq.~\eqref{eq:TEP1} and the conventional entropy
    production $\Delta {S}^{\mathrm{vN}}(t)$ in
    Eq.~\eqref{eq:EPpre3}.
    All the parameters except $N$ are the same as
    those in Fig.~\ref{fig:elements1}.}
    \label{fig:EP1}
\end{figure}
\begin{figure}
  \centering
    \includegraphics[width=0.45\textwidth]{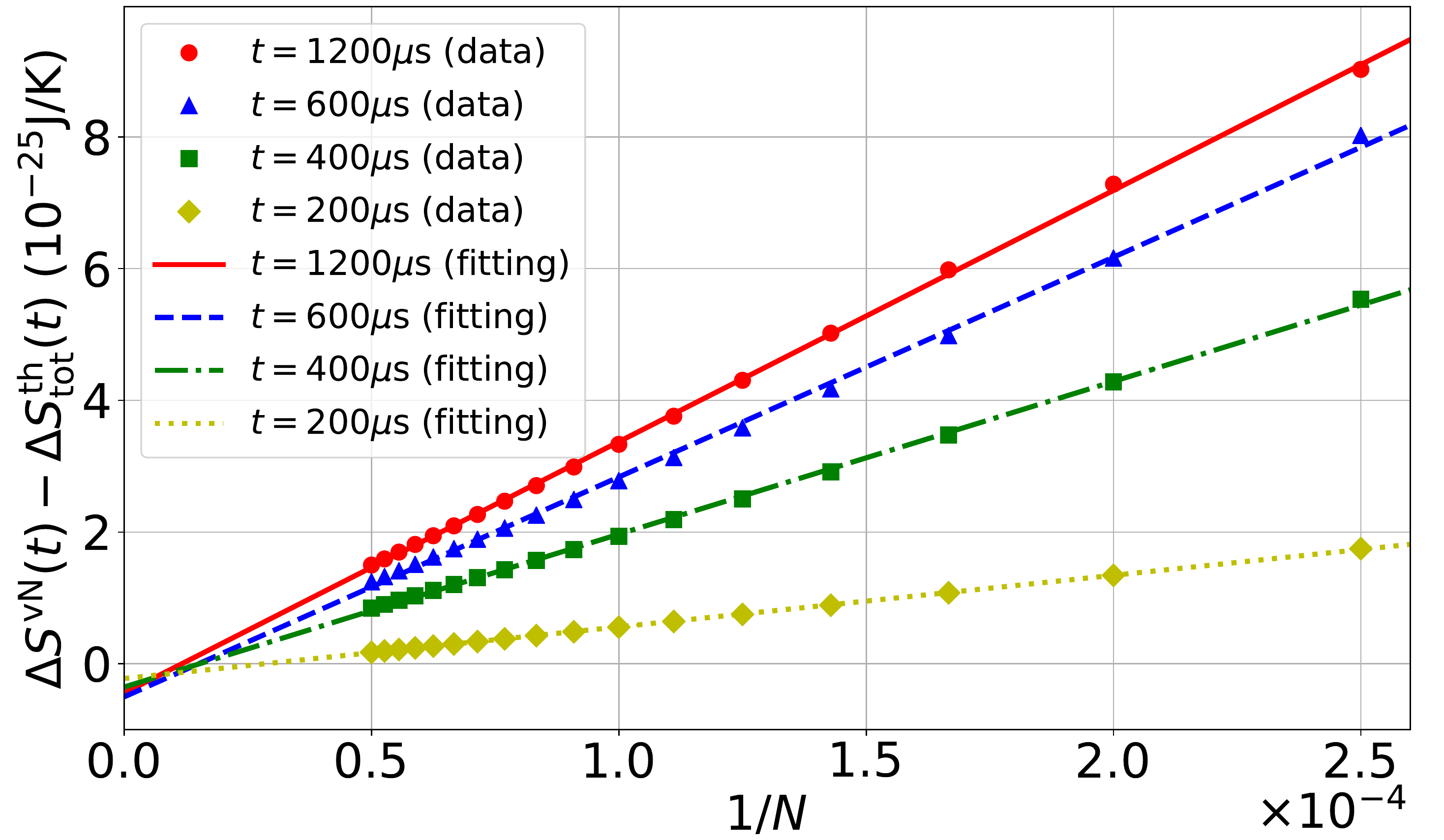}
    \caption{The difference
    $\Delta S^{\mathrm{vN}}(t)-\Delta S_{\mathrm{tot}}^{\mathrm{th}}(t)$
    in Eq.~\eqref{eq:difStot1}
    against $N^{-1}$ for four different times.
    All the parameters except $N$ are the same as
    those in Fig.~\ref{fig:elements1}.}
    \label{fig:approach1}
\end{figure}

We compare in Fig.~\ref{fig:EP1} our total thermodynamic entropy
production $\Delta S_{\mathrm{tot}}^{\mathrm{th}}(t)$
in Eq.~\eqref{eq:TEP1} with
the conventional entropy production,
which in our settings is given by
\begin{align}
  \Delta {S}^{\mathrm{vN}}(t):&=\int_0^t
  \mathrm{d}t\,{\Pi}^{\mathrm{vN}}(t)
  \notag \\
  &=S^{\mathrm{th}}_A(t)-S^{\mathrm{th}}_A(0)
  -\frac{E_A(t)-E_A(0)}{T_B^0}.
  \label{eq:EPpre3}
\end{align}
As the entropy production is the time integral
of the entropy production rate,
our total thermodynamic entropy
production approaches the conventional
entropy production as $N$ becomes larger,
which is similar to the case of the total entropy production rate.
In fact, the difference
\begin{align}
    \Delta {S}^{\mathrm{vN}}(t)-\Delta S_{\mathrm{tot}}^{\mathrm{th}}(t)
    &=\frac{E_A(0)-E_A(t)}{T_B^0}
    \notag \\
    &\quad\mbox{}+\sum_{j=2}^{N+1}\left[
    S_j^{\mathrm{th}}(0)-S_j^{\mathrm{th}}(t)
    \right]
    \label{eq:difStot1}
\end{align}
is almost proportional to $N^{-1}$ for large $N$;
see Fig.~\ref{fig:approach1}.
This suggests that $\Delta S_{\mathrm{tot}}^{\mathrm{th}}(t)$
may converge to $\Delta {S}^{\mathrm{vN}}(t)$
in the limit $N\to\infty$.

Our total thermodynamic entropy production
changes little 
for $800\,\mu$s $\lesssim t\leq1200\,\mu$s,
as shown in Fig.~\ref{fig:EP1}.
We therefore regard
the quantum state of the total system
$\hat{\rho}(t)$ in this time range
as an equilibrium state.
Since $\Delta S_{\mathrm{tot}}^{\mathrm{th}}(t)>0$
for $800\,\mu$s $\lesssim t\leq1200\,\mu$s,
we judge that our total thermodynamic entropy
$S_{\mathrm{tot}}^{\mathrm{th}}(t)$
satisfies the principle of increasing
total thermodynamic entropy
\eqref{eq:2nd_1}.

\section{Conclusion}
\label{sec:conclusion}
In conclusion, we have defined the nonequilibrium
thermodynamic entropy for the quantum model of coupled
harmonic oscillators
in a star configuration.
We analytically confirmed that our 
total thermodynamic entropy satisfies the third law
of thermodynamics. 
We have found numerically
that our total thermodynamic entropy
production rate can be negative
even when the dynamics of the 
central harmonic oscillator (system)
is well approximated by the
GKSL-type Markovian
master equation,
while our total thermodynamic entropy
satisfies the second law of thermodynamics.

Because of the specific Hamiltonian and the special initial state in our settings, all harmonic oscillators are in Gibbs states
for all the time.
This allows us to define 
the thermodynamic entropy
of each harmonic oscillator
in the present work.
If we instead prepare a different initial state,
each harmonic oscillator will be no longer
in a Gibbs state.
Defining the nonequilibrium thermodynamic entropy
of each harmonic oscillator and of the total system
in this case can be an interesting future work.

\begin{acknowledgments}
We are grateful to Naomichi Hatano for his dedicated assistance in this study.
We also appreciate Lee Jaeha's valuable comments.
 This work was supported by Leading Initiative
for Excellent Young Researchers MEXT Japan and JST
presto (Grant No. JPMJPR1919) Japan.
\end{acknowledgments}
\appendix
\begin{widetext}
\section{Every harmonic oscillator is in a Gibbs state
with a time-dependent temperature}
\label{app:Gibbs1}
In this Appendix,
we show that every harmonic oscillator
is in a Gibbs state all the time.
As each harmonic oscillator is in a
single-mode Gaussian state with
vanishing first moments,
its density operator is totally
determined by its covariance matrix \eqref{eq:sigmajt1}.
Since the time evolution of the covariance matrix
is easier to calculate than that of the density operator,
we first calculate the covariance matrix
of each harmonic oscillator at time $t$
in the next two paragraphs.
Then, in the last paragraph,
using the relation between the density
operator and the covariance matrix in Eq.~\eqref{eq:rhojt1},
we show that each harmonic oscillator is in
a Gibbs state with a time-dependent
temperature $T_j(t)$.

The matrix $H$ in Eq.~\eqref{eq:totH1}
with the elements \eqref{eq:Helements1} has a form
of the following symmetric block matrix:
\begin{align}
  H=
  \begin{pmatrix}
    \omega_1I_2 & g_2I_2 & g_3I_2 & \cdots & g_{N+1}I_2 \\
    g_2I_2 & \omega_2I_2 & 0 & \cdots & 0 \\
    g_3I_2 & 0 & \omega_3I_2 & \ddots & \vdots \\
    \vdots & \vdots & \ddots & \ddots & 0 \\
    g_{N+1}I_2 & 0 & \cdots & 0 & \omega_{N+1}I_2 \\
  \end{pmatrix}.
\end{align}
Therefore, the $n$th power of $H$
has a form of the following symmetric block matrix:
\begin{align}
  H^n=
  \begin{pmatrix}
    h_{1,1}(n)I_2 & \cdots & h_{1,N+1}(n)I_2 \\
    \vdots & \ddots & \vdots \\
    h_{1,N+1}(n)I_2 & \cdots & h_{N+1,N+1}(n)I_2 \\
  \end{pmatrix},
\end{align}
whose elements satisfy
\begin{align}
  \begin{gathered}
    \left(
    H^n
    \right)_{2j-1,2k-1}
    =\left(
    H^n
    \right)_{2j,2k},
    \quad
    \left(
    H^n
    \right)_{2j-1,2k}
    =\left(
    H^n
    \right)_{2j,2k-1}
    =0
    \\
    \mathrm{for}
    \quad
    n\in\mathbb{N},
    \quad
    j,k=1,\dots,N+1.
  \end{gathered}
  \label{eq:nthH1}
\end{align}
The $(2n-1)$th and the $(2n)$th powers
of the matrix $\Omega$ in Eq.~\eqref{eq:Omega}
are given by
\begin{align}
  \Omega^{2n-1}=(-1)^{n-1}\Omega,
  \quad
  \Omega^{2n}=(-1)^{n}I_{2N+2}
  \quad
  \mathrm{for}
  \quad
  n\in\mathbb{N},
  \label{eq:nOmega1}
\end{align}
where $I_{2N+2}$ is the $(2N+2)$-dimensional
identity matrix.
The matrices $H$ and $\Omega$ commute
with each other:
\begin{align}
  H\Omega=\Omega H.
  \label{eq:HOmega1}
\end{align}
Using Eqs.~\eqref{eq:nOmega1} and
\eqref{eq:HOmega1},
we can rewrite $V(t)=\mathrm{e}^{\Omega Ht}$
in Eq.~\eqref{eq:unitary2} as
\begin{align}
  V(t)&=\mathrm{e}^{\Omega Ht}
  \notag \\
  &=\sum_{n=0}^{\infty}
  \frac{
  \Omega^{2n}
  \left(
  Ht
  \right)^{2n}}
  {(2n)!}
  +\sum_{n=1}^{\infty}
  \frac{
  \Omega^{2n-1}
  \left(
  Ht
  \right)^{2n-1}}
  {(2n-1)!}
  \notag \\
  &=\sum_{n=0}^{\infty}
  \frac{
  (-1)^{n}
  \left(
  Ht
  \right)^{2n}}
  {(2n)!}
  +\Omega\sum_{n=1}^{\infty}
  \frac{
  (-1)^{n-1}
  \left(
  Ht
  \right)^{2n-1}}
  {(2n-1)!}
  \notag \\
  &=\cos{Ht}+\Omega\sin{Ht},
  \label{eq:St1}
\end{align}
whose transpose is
\begin{align}
  V(t)^{\mathrm{T}}=
  \cos{Ht}-[\sin{Ht}]\Omega
\end{align}
because $\Omega^{\mathrm{T}}=-\Omega$.
From Eq.~\eqref{eq:nthH1},
we find
\begin{align}
  \begin{gathered}
    \left(
    \cos{Ht}
    \right)_{2j-1,2k-1}
    =\left(
    \cos{Ht}
    \right)_{2j,2k},
    \quad
    \left(
    \cos{Ht}
    \right)_{2j-1,2k}
    =\left(
    \cos{Ht}
    \right)_{2j,2k-1}
    =0,
    \\
    \left(
    \sin{Ht}
    \right)_{2j-1,2k-1}
    =\left(
    \sin{Ht}
    \right)_{2j,2k},
    \quad
    \left(
    \sin{Ht}
    \right)_{2j-1,2k}
    =\left(
    \sin{Ht}
    \right)_{2j,2k-1}
    =0
    \\
    \mathrm{for}
    \quad
    j,k=1,\dots,N+1.
  \end{gathered}
  \label{eq:cossin1}
\end{align}

As the initial covariance matrix \eqref{eq:inicov1} is diagonal,
each element of $\sigma(t)=V(t)\sigma(0)V(t)^{\mathrm{T}}$
is written as
\begin{align}
  \begin{gathered}
    \sigma_{j,k}(t)=
    \sum_{l=1}^{2N+2}
    \left(
    \cos{Ht}+\Omega\sin{Ht}
    \right)_{j,l}
    \sigma_{l,l}(0)
    \left(
    \cos{Ht}-[\sin{Ht}]\Omega
    \right)_{l,k}
    \\
    \mathrm{for}
    \quad
    j,k=1,\dots,2N+2.
  \end{gathered}
\end{align}
Let us calculate the elements
of the covariance matrix of the $j$th
harmonic oscillator \eqref{eq:sigmajt1}.
We first obtain
\begin{align}
  \sigma_{2j-1,2j-1}(t)
  &=\sum_{l=1}^{2N+2}
  \left(
  \cos{Ht}+\Omega\sin{Ht}
  \right)_{2j-1,l}
  \sigma_{l,l}(0)
  \left(
  \cos{Ht}-[\sin{Ht}]\Omega
  \right)_{l,2j-1}
  \notag \\
  &=\sum_{l=1}^{2N+2}
  \left[
  \left(
  \cos{Ht}
  \right)_{2j-1,l}
  +\left(
  \sin{Ht}
  \right)_{2j,l}
  \right]
  \sigma_{l,l}(0)
  \left[
  \left(
  \cos{Ht}
  \right)_{l,2j-1}
  +\left(
  \sin{Ht}
  \right)_{l,2j}
  \right]
  \notag \\
  &=\sum_{m=1}^{N+1}
  \left[
  \left(
  \cos{Ht}
  \right)_{2j-1,2m-1}
  \sigma_{2m-1,2m-1}(0)
  \left(
  \cos{Ht}
  \right)_{2m-1,2j-1}
  \right.
  \notag \\
  &\quad\mbox{}+
  \left.
  \left(
  \sin{Ht}
  \right)_{2j,2m}
  \sigma_{2m,2m}(0)
  \left(
  \sin{Ht}
  \right)_{2m,2j}
  \right]
  \notag \\
  &=\sum_{m=1}^{N+1}
  \left[
  \left(
  \cos{Ht}
  \right)_{2j-1,2m-1}
  \sigma_{2m-1,2m-1}(0)
  \left(
  \cos{Ht}
  \right)_{2m-1,2j-1}
  \right.
  \notag \\
  &\quad\mbox{}+
  \left.
  \left(
  \sin{Ht}
  \right)_{2j-1,2m-1}
  \sigma_{2m-1,2m-1}(0)
  \left(
  \sin{Ht}
  \right)_{2m-1,2j-1}
  \right],
\end{align}
where the second line follows from
the form of $\Omega$ in Eq.~\eqref{eq:Omega},
the third line follows from
Eq.~\eqref{eq:cossin1},
and the last line follows
from Eq.~\eqref{eq:cossin1} and
the form of $\sigma(0)$ in Eq.~\eqref{eq:inicov1}.
Similarly, we have
\begin{align}
  \sigma_{2j,2j}(t)
  &=\sum_{l=1}^{2N+2}
  \left(
  \cos{Ht}+\Omega\sin{Ht}
  \right)_{2j,l}
  \sigma_{l,l}(0)
  \left(
  \cos{Ht}-[\sin{Ht}]\Omega
  \right)_{l,2j}
  \notag \\
  &=\sum_{l=1}^{2N+2}
  \left[
  \left(
  \cos{Ht}
  \right)_{2j,l}
  -\left(
  \sin{Ht}
  \right)_{2j-1,l}
  \right]
  \sigma_{l,l}(0)
  \left[
  \left(
  \cos{Ht}
  \right)_{l,2j}
  -\left(
  \sin{Ht}
  \right)_{l,2j-1}
  \right]
  \notag \\
  &=\sum_{m=1}^{N+1}
  \left[
  \left(
  \cos{Ht}
  \right)_{2j,2m}
  \sigma_{2m,2m}(0)
  \left(
  \cos{Ht}
  \right)_{2m,2j}
  \right.
  \notag \\
  &\quad\mbox{}+
  \left.
  \left(
  \sin{Ht}
  \right)_{2j-1,2m-1}
  \sigma_{2m-1,2m-1}(0)
  \left(
  \sin{Ht}
  \right)_{2m-1,2j-1}
  \right]
  \notag \\
  &=\sum_{m=1}^{N+1}
  \left[
  \left(
  \cos{Ht}
  \right)_{2j-1,2m-1}
  \sigma_{2m-1,2m-1}(0)
  \left(
  \cos{Ht}
  \right)_{2m-1,2j-1}
  \right.
  \notag \\
  &\quad\mbox{}+
  \left.
  \left(
  \sin{Ht}
  \right)_{2j-1,2m-1}
  \sigma_{2m-1,2m-1}(0)
  \left(
  \sin{Ht}
  \right)_{2m-1,2j-1}
  \right]
  \notag \\
  &=\sigma_{2j-1,2j-1}(t),
  \end{align}
  \begin{align}
  \sigma_{2j-1,2j}(t)
  &=\sum_{l=1}^{2N+2}
  \left(
  \cos{Ht}+\Omega\sin{Ht}
  \right)_{2j-1,l}
  \sigma_{l,l}(0)
  \left(
  \cos{Ht}-[\sin{Ht}]\Omega
  \right)_{l,2j}
  \notag \\
  &=\sum_{l=1}^{2N+2}
  \left[
  \left(
  \cos{Ht}
  \right)_{2j-1,l}
  +\left(
  \sin{Ht}
  \right)_{2j,l}
  \right]
  \sigma_{l,l}(0)
  \left[
  \left(
  \cos{Ht}
  \right)_{l,2j}
  -\left(
  \sin{Ht}
  \right)_{l,2j-1}
  \right]
  \notag \\
  &=\sum_{m=1}^{N+1}
  \left[-
  \left(
  \cos{Ht}
  \right)_{2j-1,2m-1}
  \sigma_{2m-1,2m-1}(0)
  \left(
  \sin{Ht}
  \right)_{2m-1,2j-1}
  \right.
  \notag \\
  &\quad\mbox{}+
  \left.
  \left(
  \sin{Ht}
  \right)_{2j,2m}
  \sigma_{2m,2m}(0)
  \left(
  \cos{Ht}
  \right)_{2m,2j}
  \right]
  \notag \\
  &=\sum_{m=1}^{N+1}
  \left[-
  \left(
  \cos{Ht}
  \right)_{2j-1,2m-1}
  \sigma_{2m-1,2m-1}(0)
  \left(
  \sin{Ht}
  \right)_{2m-1,2j-1}
  \right.
  \notag \\
  &\quad\mbox{}+
  \left.
  \left(
  \sin{Ht}
  \right)_{2j-1,2m-1}
  \sigma_{2m-1,2m-1}(0)
  \left(
  \cos{Ht}
  \right)_{2m-1,2j-1}
  \right]
  \notag \\
  &=\sum_{m=1}^{N+1}
  \left[-
  \left(
  \cos{Ht}
  \right)_{2j-1,2m-1}
  \sigma_{2m-1,2m-1}(0)
  \left(
  \sin{Ht}
  \right)_{2m-1,2j-1}
  \right.
  \notag \\
  &\quad\mbox{}+
  \left.
  \left(
  \sin{Ht}
  \right)_{2m-1,2j-1}
  \sigma_{2m-1,2m-1}(0)
  \left(
  \cos{Ht}
  \right)_{2j-1,2m-1}
  \right]
  \notag \\
  &=0,
\end{align}
where the fifth line follows
from the symmetry of
$\cos{Ht}$ and $\sin{Ht}$.
We thus arrive at
\begin{align}
  \sigma_j(t)=\sigma_{2j-1,2j-1}(t)I_2,
  \label{eq:sigmajt3}
\end{align}
which appears in Eq.~\eqref{eq:sigmajt2}
in the main text.

As the $j$th harmonic oscillator
is in a single-mode Gaussian state
with vanishing first moments,
its density operator is
completely characterized
by the covariance matrix in Eq.~\eqref{eq:sigmajt3}
and has the following form
\cite{PhysRevLett.115.260501}:
\begin{align}
  \hat{\rho}_j(t)
  =\frac{\exp\left[
  -\frac{1}{2}\hat{\bf{r}}_j^{\mathrm{T}}
  G_j(t)\hat{\bf{r}}_j
  \right]}
  {Z_j(t)},
  \label{eq:rhojt1}
\end{align}
where
\begin{gather}
  \hat{\bf{r}}_j=\left(
  \hat{r}_{2j-1},\hat{r}_{2j}
  \right)^{\mathrm{T}},
  \label{eq:rj1} \\
  G_j(t)=2\mathrm{i}\Omega_1
  \coth^{-1}\left[
  \sigma_j(t)\mathrm{i}\Omega_1
  \right]
  =2\coth^{-1}\left[
  \sigma_{2j-1,2j-1}(t)
  \right]I_2
  \quad\mathrm{with}\quad
  \Omega_1=
  \begin{pmatrix}
    0 & 1 \\
    -1 & 0 \\
  \end{pmatrix},
  \label{eq:Gjt1} \\
  Z_j(t)=\frac{1}{2}\sqrt{
  \det\left(
  \sigma_j(t)+\mathrm{i}\Omega_1
  \right)
  }
  =\frac{1}{2}\sqrt{
  \sigma_{2j-1,2j-1}(t)^2-1
  }.
\end{gather}
Let us show that $\hat{\rho}_j(t)$ in Eq.~\eqref{eq:rhojt1}
is a Gibbs state below.
The numerator of Eq.~\eqref{eq:rhojt1} is transformed using
Eqs.~\eqref{eq:rj1} and \eqref{eq:Gjt1} as
\begin{align}
  \exp\left[
  -\frac{1}{2}\hat{\bf{r}}_j^{\mathrm{T}}
  G_j(t)\hat{\bf{r}}_j
  \right]
  =\exp\left[
  -\frac{2}{\hbar\omega_j}
  \coth^{-1}\left[
  \sigma_{2j-1,2j-1}(t)
  \right]
  \frac{\hbar\omega_j}{2}
  \left(
  \hat{r}_{2j-1}^2+\hat{r}_{2j}^2
  \right)
  \right]
  =\exp\left[
  -\beta_j(t)\hat{H}_j
  \right],
  \label{eq:exp1}
\end{align}
where
\begin{align}
  \beta_j(t)=\frac{1}{k_BT_j(t)}
  =\frac{2}{\hbar\omega_j}
  \coth^{-1}\left[
  \sigma_{2j-1,2j-1}(t)
  \right]
  =\frac{1}{\hbar\omega_j}
  \ln\left(
  \frac{\sigma_{2j-1,2j-1}(t)+1}
  {\sigma_{2j-1,2j-1}(t)-1}
  \right).
  \label{eq:betajt2}
\end{align}
The trace of the numerator of Eq.~\eqref{eq:rhojt1}
is equal to the denominator:
\begin{align}
  \mathrm{Tr}\left[
  \exp\left[
  -\frac{1}{2}\hat{\bf{r}}_j^{\mathrm{T}}
  G_j(t)\hat{\bf{r}}_j
  \right]
  \right]
  &=
  \mathrm{Tr}\left[
  \exp\left[
  -\beta_j(t)\hat{H}_j
  \right]
  \right]
  \notag \\
  &=\mathrm{Tr}\left[
  \exp\left[
  -\beta_j(t)\hbar\omega_j
  \left(\hat{a}_j^{\dag}\hat{a}_j+\frac{1}{2}
  \right)
  \right]
  \right]
  \notag \\
  &=\sum_{n=0}^{\infty}
  \exp\left[
  -\beta_j(t)\hbar\omega_j
  \left(n+\frac{1}{2}
  \right)
  \right]
  \notag \\
  &=\frac{\exp\left[
  -\beta_j(t)\hbar\omega_j/2
  \right]}
  {1-\exp\left[
  -\beta_j(t)\hbar\omega_j
  \right]}
  \notag \\
  &=\frac{1}
  {\exp\left[
  \beta_j(t)\hbar\omega_j/2
  \right]-\exp\left[
  -\beta_j(t)\hbar\omega_j/2
  \right]}
  \notag \\
  &=\frac{1}{
  \sqrt{\frac{\sigma_{2j-1,2j-1}(t)+1}
  {\sigma_{2j-1,2j-1}(t)-1}}
  -\sqrt{\frac{\sigma_{2j-1,2j-1}(t)-1}
  {\sigma_{2j-1,2j-1}(t)+1}}
  }
  \notag \\
  &=\frac{1}{2}\sqrt{
  \sigma_{2j-1,2j-1}(t)^2-1
  }
  \notag \\
  &=Z_j(t),
\end{align}
where in the sixth line, we have used Eq.~\eqref{eq:betajt2}.
Therefore, we have derived Eqs.~\eqref{eq:rhojt2}
and \eqref{eq:Zjt2}:
\begin{align}
  \hat{\rho}_j(t)=
  \frac{\mathrm{e}^{-\beta_j(t)\hat{H}_j}}
  {Z_j(t)},\quad
  Z_j(t)=\mathrm{Tr}\left[
  \mathrm{e}^{-\beta_j(t)\hat{H}_j}
  \right]
  =\frac{1}{2}\sqrt{
  \sigma_{2j-1,2j-1}(t)^2-1
  },
\end{align}
which shows that the $j$th harmonic oscillator
is in the Gibbs state with the time-dependent
inverse temperature $\beta_j(t)$.

\section{The time derivative of the mean energy
of each harmonic oscillator,
the bath, and the interaction}
\label{app:dEdt}
In this Appendix we calculate the time derivative
of the mean energy of each harmonic oscillator
so that we can transform the thermodynamic
entropy production rate \eqref{eq:TEPR1}
to the more easily calculable form \eqref{eq:TEPR2}.
We also calculate the time derivative
of the mean energy of the bath and the interaction
in order to show in Fig.~\ref{fig:EABIt1}
that the interaction energy is negligibly small.

The modified position and momentum
operators introduced in Eq.~\eqref{eq:modified}
satisfy the canonical commutation
relations:
\begin{align}
  \left[\hat{r}_j,\hat{r}_k\right]
  =\mathrm{i}\Omega_{j,k}
  \quad\mathrm{for}\quad j,k=1,\dots,2(N+1),
  \label{eq:CCR1}
\end{align}
The total Hamiltonian in the Heisenberg
picture is
\begin{align}
  \hat{H}^{H}(t)=\sum_{j=1}^{N+1}
  \frac{\hbar\omega_j}{2}
  \left(\hat{r}_{2j-1}^H(t)^2
  +\hat{r}_{2j}^H(t)^2
  \right)
  +\sum_{j=2}^{N+1}
  \hbar g_j
  \left(\hat{r}_1^H(t)\hat{r}_{2j-1}^H(t)
  +\hat{r}_2^H(t)\hat{r}_{2j}^H(t)
  \right),
\end{align}
where $\hat{r}_k^H(t)=\hat{U}^{\dag}(t)\hat{r}_k\hat{U}(t)$
for $k=1,\dots,N+1$.
The Heisenberg equations of motions read
\begin{gather}
  \frac{\mathrm{d}}{\mathrm{d}t}
  \hat{r}_1^H(t)
  =\frac{\mathrm{i}}{\hbar}
  \left[\hat{H}^H(t),\hat{r}_1^H(t)\right]
  =\omega_1\hat{r}_2^H(t)
  +\sum_{j=2}^{N+1}g_j\hat{r}_{2j}^H(t),
  \\
  \frac{\mathrm{d}}{\mathrm{d}t}
  \hat{r}_2^H(t)
  =\frac{\mathrm{i}}{\hbar}
  \left[\hat{H}^H(t),\hat{r}_2^H(t)\right]
  =-\omega_1\hat{r}_1^H(t)
  -\sum_{j=2}^{N+1}g_j\hat{r}_{2j-1}^H(t),
  \\
  \frac{\mathrm{d}}{\mathrm{d}t}
  \hat{r}_{2j-1}^H(t)
  =\frac{\mathrm{i}}{\hbar}
  \left[\hat{H}^H(t),\hat{r}_{2j-1}^H(t)\right]
  =\omega_j\hat{r}_{2j}^H(t)
  +g_j\hat{r}_{2}^H(t)
  \quad\mathrm{for}\quad j=2,\dots,N+1,
  \\
  \frac{\mathrm{d}}{\mathrm{d}t}
  \hat{r}_{2j}^H(t)
  =\frac{\mathrm{i}}{\hbar}
  \left[\hat{H}^H(t),\hat{r}_{2j}^H(t)\right]
  =-\omega_j\hat{r}_{2j-1}^H(t)
  -g_j\hat{r}_{1}^H(t)
  \quad\mathrm{for}\quad j=2,\dots,N+1.
\end{gather}

Then, the time derivative of the mean energy
of each harmonic oscillator is calculated as
\begin{align}
  \frac{\mathrm{d}}{\mathrm{d}t}
  E_A(t)
  &=\frac{\mathrm{d}}{\mathrm{d}t}
  E_1(t)
  \notag \\
  &=\frac{\hbar\omega_1}{2}
  \frac{\mathrm{d}}{\mathrm{d}t}
  \sigma_{1,1}(t)
  \notag \\
  &=\frac{\hbar\omega_1}{2}
  \frac{\mathrm{d}}{\mathrm{d}t}
  \mathrm{Tr}
  \left[
  \hat{\rho}(0)
  \left\{
  \hat{r}_1^H(t),\hat{r}_1^H(t)
  \right\}
  \right]
  \notag \\
  &=\frac{\hbar\omega_1}{2}
  \mathrm{Tr}
  \left[
  \hat{\rho}(0)
  \left\{
  \frac{\mathrm{d}}{\mathrm{d}t}
  \hat{r}_1^H(t),\hat{r}_1^H(t)
  \right\}
  \right]
  +\frac{\hbar\omega_1}{2}
  \mathrm{Tr}
  \left[
  \hat{\rho}(0)
  \left\{
  \hat{r}_1^H(t),
  \frac{\mathrm{d}}{\mathrm{d}t}
  \hat{r}_1^H(t)
  \right\}
  \right]
  \notag \\
  &=\frac{\hbar\omega_1}{2}
  \mathrm{Tr}
  \left[
  \hat{\rho}(0)
  \left\{
  \left(
  \omega_1\hat{r}_2^H(t)
  +\sum_{j=2}^{N+1}g_j\hat{r}_{2j}^H(t)
  \right),
  \hat{r}_1^H(t)
  \right\}
  \right]
  \notag \\
  &\quad
  \mbox{}
  +\frac{\hbar\omega_1}{2}
  \mathrm{Tr}
  \left[
  \hat{\rho}(0)
  \left\{
  \hat{r}_1^H(t),
  \left(
  \omega_1\hat{r}_2^H(t)
  +\sum_{j=2}^{N+1}g_j\hat{r}_{2j}^H(t)
  \right)
  \right\}
  \right]
  \notag \\
  &=\frac{\hbar\omega_1}{2}
  \left[
  \omega_1\sigma_{2,1}(t)
  +\sum_{j=2}^{N+1}g_j
  \sigma_{2j,1}(t)
  \right]
  +\frac{\hbar\omega_1}{2}
  \left[
  \omega_1\sigma_{1,2}(t)
  +\sum_{j=2}^{N+1}g_j
  \sigma_{1,2j}(t)
  \right]
  \notag \\
  &=\hbar\omega_1
  \left[
  \omega_1\sigma_{1,2}(t)
  +\sum_{j=2}^{N+1}g_j
  \sigma_{1,2j}(t)
  \right]
  \notag \\
  &=\hbar\omega_1
  \sum_{j=2}^{N+1}g_j
  \sigma_{1,2j}(t),
  \label{eq:dE1tdt1}
  \end{align}
\begin{align}
  \frac{\mathrm{d}}{\mathrm{d}t}
  E_j(t)
  &=\frac{\hbar\omega_j}{2}
  \frac{\mathrm{d}}{\mathrm{d}t}
  \sigma_{2j-1,2j-1}(t)
  \notag \\
  &=\frac{\hbar\omega_j}{2}
  \frac{\mathrm{d}}{\mathrm{d}t}
  \mathrm{Tr}
  \left[
  \hat{\rho}(0)
  \left\{
  \hat{r}_{2j-1}^H(t),\hat{r}_{2j-1}^H(t)
  \right\}
  \right]
  \notag \\
  &=\frac{\hbar\omega_j}{2}
  \mathrm{Tr}
  \left[
  \hat{\rho}(0)
  \left\{
  \frac{\mathrm{d}}{\mathrm{d}t}
  \hat{r}_{2j-1}^H(t),\hat{r}_{2j-1}^H(t)
  \right\}
  \right]
  +\frac{\hbar\omega_j}{2}
  \mathrm{Tr}
  \left[
  \hat{\rho}(0)
  \left\{
  \hat{r}_{2j-1}^H(t),
  \frac{\mathrm{d}}{\mathrm{d}t}
  \hat{r}_{2j-1}^H(t)
  \right\}
  \right]
  \notag \\
  &=\frac{\hbar\omega_j}{2}
  \mathrm{Tr}
  \left[
  \hat{\rho}(0)
  \left\{
  \left(
  \omega_j\hat{r}_{2j}^H(t)
  +g_j\hat{r}_{2}^H(t)
  \right),\hat{r}_{2j-1}^H(t)
  \right\}
  \right]
  +\frac{\hbar\omega_j}{2}
  \mathrm{Tr}
  \left[
  \hat{\rho}(0)
  \left\{
  \hat{r}_{2j-1}^H(t),
  \left(
  \omega_j\hat{r}_{2j}^H(t)
  +g_j\hat{r}_{2}^H(t)
  \right)
  \right\}
  \right]
  \notag \\
  &=\frac{\hbar\omega_j}{2}
  \left[
  \omega_j\sigma_{2j,2j-1}(t)
  +g_j\sigma_{2,2j-1}(t)
  \right]
  +\frac{\hbar\omega_j}{2}
  \left[
  \omega_j\sigma_{2j-1,2j}(t)
  +g_j\sigma_{2j-1,2}(t)
  \right]
  \notag \\
  &=\hbar\omega_j
  \left[
  \omega_j\sigma_{2j-1,2j}(t)
  +g_j\sigma_{2,2j-1}(t)
  \right]
  \notag \\
  &=\hbar\omega_j
  g_j\sigma_{2,2j-1}(t)
  \notag \\
  &=-\hbar\omega_j
  g_j\sigma_{1,2j}(t)
  \quad\mathrm{for}\quad j=2,\dots,N+1,
  \label{eq:dEjtdt1}
\end{align}
where the last line follows from
\begin{align}
  \sigma_{1,2j}(t)
  &=\sum_{l=1}^{2N+2}
  \left(
  \cos{Ht}+\Omega\sin{Ht}
  \right)_{1,l}
  \sigma_{l,l}(0)
  \left(
  \cos{Ht}-[\sin{Ht}]\Omega
  \right)_{l,2j}
  \notag \\
  &=\sum_{l=1}^{2N+2}
  \left[
  \left(
  \cos{Ht}
  \right)_{1,l}
  +\left(
  \sin{Ht}
  \right)_{2,l}
  \right]
  \sigma_{l,l}(0)
  \left[
  \left(
  \cos{Ht}
  \right)_{l,2j}
  -\left(
  \sin{Ht}
  \right)_{l,2j-1}
  \right]
  \notag \\
  &=\sum_{m=1}^{N+1}
  \left[
  -\left(
  \cos{Ht}
  \right)_{1,2m-1}
  \sigma_{2m-1,2m-1}(0)
  \left(
  \sin{Ht}
  \right)_{2m-1,2j-1}
  +\left(
  \sin{Ht}
  \right)_{2,2m}
  \sigma_{2m,2m}(0)
  \left(
  \cos{Ht}
  \right)_{2m,2j}
  \right]
  \notag \\
  &=\sum_{m=1}^{N+1}
  \left[
  -\left(
  \cos{Ht}
  \right)_{1,2m-1}
  \sigma_{2m-1,2m-1}(0)
  \left(
  \sin{Ht}
  \right)_{2m-1,2j-1}
  \right.
  \notag \\
  &\quad\mbox{}+
  \left.
  \left(
  \sin{Ht}
  \right)_{1,2m-1}
  \sigma_{2m-1,2m-1}(0)
  \left(
  \cos{Ht}
  \right)_{2m-1,2j-1}
  \right],
  \end{align}
\begin{align}
  \sigma_{2,2j-1}(t)
  &=\sum_{l=1}^{2N+2}
  \left(
  \cos{Ht}+\Omega\sin{Ht}
  \right)_{2,l}
  \sigma_{l,l}(0)
  \left(
  \cos{Ht}-[\sin{Ht}]\Omega
  \right)_{l,2j-1}
  \notag \\
  &=\sum_{l=1}^{2N+2}
  \left[
  \left(
  \cos{Ht}
  \right)_{2,l}
  -\left(
  \sin{Ht}
  \right)_{1,l}
  \right]
  \sigma_{l,l}(0)
  \left[
  \left(
  \cos{Ht}
  \right)_{l,2j-1}
  +\left(
  \sin{Ht}
  \right)_{l,2j}
  \right]
  \notag \\
  &=\sum_{m=1}^{N+1}
  \left[
  \left(
  \cos{Ht}
  \right)_{2,2m}
  \sigma_{2m,2m}(0)
  \left(
  \sin{Ht}
  \right)_{2m,2j}
  -\left(
  \sin{Ht}
  \right)_{1,2m-1}
  \sigma_{2m-1,2m-1}(0)
  \left(
  \cos{Ht}
  \right)_{2m-1,2j-1}
  \right]
  \notag \\
  &=\sum_{m=1}^{N+1}
  \left[
  \left(
  \cos{Ht}
  \right)_{1,2m-1}
  \sigma_{2m-1,2m-1}(0)
  \left(
  \sin{Ht}
  \right)_{2m-1,2j-1}
  \right.
  \notag \\
  &\quad\mbox{}-
  \left.
  \left(
  \sin{Ht}
  \right)_{1,2m-1}
  \sigma_{2m-1,2m-1}(0)
  \left(
  \cos{Ht}
  \right)_{2m-1,2j-1}
  \right]
  \notag \\
  &=-\sigma_{1,2j}(t).
\end{align}
Inserting Eqs.~\eqref{eq:dE1tdt1} and \eqref{eq:dEjtdt1}
into Eq.~\eqref{eq:TEPR1}, we obtain Eq.~\eqref{eq:TEPR2},
which we can calculate from the covariance matrix
$\sigma(t)$.

The time derivative of the mean energy
of the bath is calculated as
\begin{align}
    \frac{\mathrm{d}}{\mathrm{d}t}
    E_B(t)=
    \sum_{j=2}^{N+1}
    \frac{\mathrm{d}}{\mathrm{d}t}
    E_j(t)
    =-\sum_{j=2}^{N+1}
    \hbar\omega_j
    g_j\sigma_{1,2j}(t).
    \label{eq:dEBtdt1}
\end{align}
From the conservation of the total energy,
the time derivative of the interaction energy
is calculated as
\begin{align}
    \frac{\mathrm{d}}{\mathrm{d}t}E_I(t)=
    -\frac{\mathrm{d}}{\mathrm{d}t}E_A(t)
    -\frac{\mathrm{d}}{\mathrm{d}t}E_B(t)
    =\sum_{j=2}^{N+1}
    \hbar(\omega_j-\omega_1)
    g_j\sigma_{1,2j}(t).
    \label{eq:dEItdt1}
\end{align}
In Fig.~\ref{fig:EABIt1}, we compare
$\mathrm{d}E_A(t)/\mathrm{d}t$
in Eq.~\eqref{eq:dE1tdt1},
$\mathrm{d}E_B(t)/\mathrm{d}t$ in
Eq.~\eqref{eq:dEBtdt1}, and
$\mathrm{d}E_I(t)/\mathrm{d}t$ in
Eq.~\eqref{eq:dEItdt1}.
\end{widetext}
\bibliography{1ref}

\begin{thebibliography}{44}%
\makeatletter
\providecommand \@ifxundefined [1]{%
 \@ifx{#1\undefined}
}%
\providecommand \@ifnum [1]{%
 \ifnum #1\expandafter \@firstoftwo
 \else \expandafter \@secondoftwo
 \fi
}%
\providecommand \@ifx [1]{%
 \ifx #1\expandafter \@firstoftwo
 \else \expandafter \@secondoftwo
 \fi
}%
\providecommand \natexlab [1]{#1}%
\providecommand \enquote  [1]{``#1''}%
\providecommand \bibnamefont  [1]{#1}%
\providecommand \bibfnamefont [1]{#1}%
\providecommand \citenamefont [1]{#1}%
\providecommand \href@noop [0]{\@secondoftwo}%
\providecommand \href [0]{\begingroup \@sanitize@url \@href}%
\providecommand \@href[1]{\@@startlink{#1}\@@href}%
\providecommand \@@href[1]{\endgroup#1\@@endlink}%
\providecommand \@sanitize@url [0]{\catcode `\\12\catcode `\$12\catcode
  `\&12\catcode `\#12\catcode `\^12\catcode `\_12\catcode `\%12\relax}%
\providecommand \@@startlink[1]{}%
\providecommand \@@endlink[0]{}%
\providecommand \url  [0]{\begingroup\@sanitize@url \@url }%
\providecommand \@url [1]{\endgroup\@href {#1}{\urlprefix }}%
\providecommand \urlprefix  [0]{URL }%
\providecommand \Eprint [0]{\href }%
\providecommand \doibase [0]{http://dx.doi.org/}%
\providecommand \selectlanguage [0]{\@gobble}%
\providecommand \bibinfo  [0]{\@secondoftwo}%
\providecommand \bibfield  [0]{\@secondoftwo}%
\providecommand \translation [1]{[#1]}%
\providecommand \BibitemOpen [0]{}%
\providecommand \bibitemStop [0]{}%
\providecommand \bibitemNoStop [0]{.\EOS\space}%
\providecommand \EOS [0]{\spacefactor3000\relax}%
\providecommand \BibitemShut  [1]{\csname bibitem#1\endcsname}%
\let\auto@bib@innerbib\@empty
\bibitem [{\citenamefont {Oono}(2017)}]{oono2017perspectives}%
  \BibitemOpen
  \bibfield  {author} {\bibinfo {author} {\bibfnamefont {Yoshitsugu}\
  \bibnamefont {Oono}},\ }\href@noop {} {\emph {\bibinfo {title} {Perspectives
  on Statistical Thermodynamics}}}\ (\bibinfo  {publisher} {Cambridge
  University Press},\ \bibinfo {address} {Cambridge, England},\ \bibinfo {year}
  {2017})\BibitemShut {NoStop}%
\bibitem [{\citenamefont {Callen}(1985)}]{callen1985thermodynamics}%
  \BibitemOpen
  \bibfield  {author} {\bibinfo {author} {\bibfnamefont {Herbert~B}\
  \bibnamefont {Callen}},\ }\href@noop {} {\emph {\bibinfo {title}
  {Thermodynamics and an Introduction to Thermostatistics}}}\ (\bibinfo
  {publisher} {Wiley},\ \bibinfo {address} {New York},\ \bibinfo {year}
  {1985})\BibitemShut {NoStop}%
\bibitem [{\citenamefont {Lebon}\ \emph {et~al.}(2008)\citenamefont {Lebon},
  \citenamefont {Jou},\ and\ \citenamefont
  {Casas-V{\'a}zquez}}]{lebon2008understanding}%
  \BibitemOpen
  \bibfield  {author} {\bibinfo {author} {\bibfnamefont {Georgy}\ \bibnamefont
  {Lebon}}, \bibinfo {author} {\bibfnamefont {David}\ \bibnamefont {Jou}}, \
  and\ \bibinfo {author} {\bibfnamefont {Jos{\'e}}\ \bibnamefont
  {Casas-V{\'a}zquez}},\ }\href@noop {} {\emph {\bibinfo {title} {Understanding
  Non-equilibrium Thermodynamics}}}\ (\bibinfo  {publisher} {Springer-Verlag},\
  \bibinfo {address} {Berlin},\ \bibinfo {year} {2008})\BibitemShut {NoStop}%
\bibitem [{\citenamefont {Binder}\ \emph {et~al.}(2018)\citenamefont {Binder},
  \citenamefont {Correa}, \citenamefont {Gogolin}, \citenamefont {Anders},\
  and\ \citenamefont {Adesso}}]{binder2018thermodynamics}%
  \BibitemOpen
  \bibfield  {author} {\bibinfo {author} {\bibfnamefont {Felix}\ \bibnamefont
  {Binder}}, \bibinfo {author} {\bibfnamefont {Luis~A}\ \bibnamefont {Correa}},
  \bibinfo {author} {\bibfnamefont {Christian}\ \bibnamefont {Gogolin}},
  \bibinfo {author} {\bibfnamefont {Janet}\ \bibnamefont {Anders}}, \ and\
  \bibinfo {author} {\bibfnamefont {Gerardo}\ \bibnamefont {Adesso}},\
  }\href@noop {} {\emph {\bibinfo {title} {Thermodynamics in the quantum
  regime}}},\ \bibinfo {series} {Fundamental Theories of Physics}, Vol.\
  \bibinfo {volume} {195}\ (\bibinfo  {publisher} {Springer},\ \bibinfo
  {address} {Cham},\ \bibinfo {year} {2018})\BibitemShut {NoStop}%
\bibitem [{\citenamefont {Vinjanampathy}\ and\ \citenamefont
  {Anders}(2016)}]{doi:10.1080/00107514.2016.1201896}%
  \BibitemOpen
  \bibfield  {author} {\bibinfo {author} {\bibfnamefont {Sai}\ \bibnamefont
  {Vinjanampathy}}\ and\ \bibinfo {author} {\bibfnamefont {Janet}\ \bibnamefont
  {Anders}},\ }\bibfield  {title} {\enquote {\bibinfo {title} {Quantum
  thermodynamics},}\ }\href {\doibase 10.1080/00107514.2016.1201896} {\bibfield
   {journal} {\bibinfo  {journal} {Contemporary Physics}\ }\textbf {\bibinfo
  {volume} {57}},\ \bibinfo {pages} {545--579} (\bibinfo {year}
  {2016})}\BibitemShut {NoStop}%
\bibitem [{\citenamefont {Kosloff}(2013)}]{e15062100}%
  \BibitemOpen
  \bibfield  {author} {\bibinfo {author} {\bibfnamefont {Ronnie}\ \bibnamefont
  {Kosloff}},\ }\bibfield  {title} {\enquote {\bibinfo {title} {Quantum
  thermodynamics: A dynamical viewpoint},}\ }\href {\doibase 10.3390/e15062100}
  {\bibfield  {journal} {\bibinfo  {journal} {Entropy}\ }\textbf {\bibinfo
  {volume} {15}},\ \bibinfo {pages} {2100--2128} (\bibinfo {year}
  {2013})}\BibitemShut {NoStop}%
\bibitem [{\citenamefont {Strasberg}\ and\ \citenamefont
  {Winter}(2021)}]{strasberg2021second}%
  \BibitemOpen
  \bibfield  {author} {\bibinfo {author} {\bibfnamefont {Philipp}\ \bibnamefont
  {Strasberg}}\ and\ \bibinfo {author} {\bibfnamefont {Andreas}\ \bibnamefont
  {Winter}},\ }\href@noop {} {\enquote {\bibinfo {title} {First and second law
  of quantum thermodynamics: A consistent derivation based on a microscopic
  definition of entropy},}\ } (\bibinfo {year} {2021}),\ \Eprint
  {http://arxiv.org/abs/2002.08817} {arXiv:2002.08817 [quant-ph]} \BibitemShut
  {NoStop}%
\bibitem [{\citenamefont {\ifmmode~\check{S}\else \v{S}\fi{}afr\'anek}\ \emph
  {et~al.}(2019)\citenamefont {\ifmmode~\check{S}\else \v{S}\fi{}afr\'anek},
  \citenamefont {Deutsch},\ and\ \citenamefont {Aguirre}}]{PhysRevA.99.012103}%
  \BibitemOpen
  \bibfield  {author} {\bibinfo {author} {\bibfnamefont {Dominik}\ \bibnamefont
  {\ifmmode~\check{S}\else \v{S}\fi{}afr\'anek}}, \bibinfo {author}
  {\bibfnamefont {J.~M.}\ \bibnamefont {Deutsch}}, \ and\ \bibinfo {author}
  {\bibfnamefont {Anthony}\ \bibnamefont {Aguirre}},\ }\bibfield  {title}
  {\enquote {\bibinfo {title} {Quantum coarse-grained entropy and
  thermalization in closed systems},}\ }\href {\doibase
  10.1103/PhysRevA.99.012103} {\bibfield  {journal} {\bibinfo  {journal} {Phys.
  Rev. A}\ }\textbf {\bibinfo {volume} {99}},\ \bibinfo {pages} {012103}
  (\bibinfo {year} {2019})}\BibitemShut {NoStop}%
\bibitem [{\citenamefont {Goldstein}\ \emph {et~al.}(2020)\citenamefont
  {Goldstein}, \citenamefont {Lebowitz}, \citenamefont {Tumulka},\ and\
  \citenamefont {Zangh{\`\i}}}]{doi:10.1142/9789811211720_0014}%
  \BibitemOpen
  \bibfield  {author} {\bibinfo {author} {\bibfnamefont {Sheldon}\ \bibnamefont
  {Goldstein}}, \bibinfo {author} {\bibfnamefont {Joel~L}\ \bibnamefont
  {Lebowitz}}, \bibinfo {author} {\bibfnamefont {Roderich}\ \bibnamefont
  {Tumulka}}, \ and\ \bibinfo {author} {\bibfnamefont {Nino}\ \bibnamefont
  {Zangh{\`\i}}},\ }\enquote {\bibinfo {title} {Gibbs and boltzmann entropy in
  classical and quantum mechanics},}\ in\ \href {\doibase
  10.1142/9789811211720_0014} {\emph {\bibinfo {booktitle} {Statistical
  Mechanics and Scientific Explanation}}}\ (\bibinfo  {publisher} {World
  Scientific},\ \bibinfo {address} {Singapore},\ \bibinfo {year} {2020})\
  Chap.~\bibinfo {chapter} {14}, pp.\ \bibinfo {pages} {519--581}\BibitemShut
  {NoStop}%
\bibitem [{\citenamefont {Landi}\ and\ \citenamefont
  {Paternostro}(2020)}]{landi2020irreversible}%
  \BibitemOpen
  \bibfield  {author} {\bibinfo {author} {\bibfnamefont {Gabriel~T.}\
  \bibnamefont {Landi}}\ and\ \bibinfo {author} {\bibfnamefont {Mauro}\
  \bibnamefont {Paternostro}},\ }\href@noop {} {\enquote {\bibinfo {title}
  {Irreversible entropy production, from quantum to classical},}\ } (\bibinfo
  {year} {2020}),\ \Eprint {http://arxiv.org/abs/2009.07668} {arXiv:2009.07668
  [quant-ph]} \BibitemShut {NoStop}%
\bibitem [{\citenamefont {Goold}\ \emph {et~al.}(2016)\citenamefont {Goold},
  \citenamefont {Huber}, \citenamefont {Riera}, \citenamefont {del Rio},\ and\
  \citenamefont {Skrzypczyk}}]{Goold_2016}%
  \BibitemOpen
  \bibfield  {author} {\bibinfo {author} {\bibfnamefont {John}\ \bibnamefont
  {Goold}}, \bibinfo {author} {\bibfnamefont {Marcus}\ \bibnamefont {Huber}},
  \bibinfo {author} {\bibfnamefont {Arnau}\ \bibnamefont {Riera}}, \bibinfo
  {author} {\bibfnamefont {L{\'{\i}}dia}\ \bibnamefont {del Rio}}, \ and\
  \bibinfo {author} {\bibfnamefont {Paul}\ \bibnamefont {Skrzypczyk}},\
  }\bibfield  {title} {\enquote {\bibinfo {title} {The role of quantum
  information in thermodynamics{\textemdash}a topical review},}\ }\href
  {\doibase 10.1088/1751-8113/49/14/143001} {\bibfield  {journal} {\bibinfo
  {journal} {Journal of Physics A: Mathematical and Theoretical}\ }\textbf
  {\bibinfo {volume} {49}},\ \bibinfo {pages} {143001} (\bibinfo {year}
  {2016})}\BibitemShut {NoStop}%
\bibitem [{\citenamefont {Tan}\ \emph {et~al.}(2020)\citenamefont {Tan},
  \citenamefont {Schwonnek}, \citenamefont {Goh}, \citenamefont {Primaatmaja},\
  and\ \citenamefont {Lim}}]{tan2020computing}%
  \BibitemOpen
  \bibfield  {author} {\bibinfo {author} {\bibfnamefont {Ernest Y.~Z.}\
  \bibnamefont {Tan}}, \bibinfo {author} {\bibfnamefont {René}\ \bibnamefont
  {Schwonnek}}, \bibinfo {author} {\bibfnamefont {Koon~Tong}\ \bibnamefont
  {Goh}}, \bibinfo {author} {\bibfnamefont {Ignatius~William}\ \bibnamefont
  {Primaatmaja}}, \ and\ \bibinfo {author} {\bibfnamefont {Charles C.~W.}\
  \bibnamefont {Lim}},\ }\href@noop {} {\enquote {\bibinfo {title} {Computing
  secure key rates for quantum key distribution with untrusted devices},}\ }
  (\bibinfo {year} {2020}),\ \Eprint {http://arxiv.org/abs/1908.11372}
  {arXiv:1908.11372 [quant-ph]} \BibitemShut {NoStop}%
\bibitem [{\citenamefont {Rivas}\ and\ \citenamefont
  {Huelga}(2012)}]{rivas2012open}%
  \BibitemOpen
  \bibfield  {author} {\bibinfo {author} {\bibfnamefont {Angel}\ \bibnamefont
  {Rivas}}\ and\ \bibinfo {author} {\bibfnamefont {Susana~F}\ \bibnamefont
  {Huelga}},\ }\href@noop {} {\emph {\bibinfo {title} {Open Quantum
  Systems}}},\ \bibinfo {series} {SpringerBriefs in Physics}, Vol.~\bibinfo
  {volume} {13}\ (\bibinfo  {publisher} {Springer},\ \bibinfo {address}
  {Berlin},\ \bibinfo {year} {2012})\BibitemShut {NoStop}%
\bibitem [{\citenamefont {Breuer}\ and\ \citenamefont
  {Petruccione}(2002)}]{breuer2002theory}%
  \BibitemOpen
  \bibfield  {author} {\bibinfo {author} {\bibfnamefont {Heinz-Peter}\
  \bibnamefont {Breuer}}\ and\ \bibinfo {author} {\bibfnamefont {Francesco}\
  \bibnamefont {Petruccione}},\ }\href@noop {} {\emph {\bibinfo {title} {The
  Theory of Open Quantum Systems}}}\ (\bibinfo  {publisher} {Oxford University
  Press},\ \bibinfo {address} {Oxford, England},\ \bibinfo {year}
  {2002})\BibitemShut {NoStop}%
\bibitem [{\citenamefont {Marcantoni}\ \emph {et~al.}(2017)\citenamefont
  {Marcantoni}, \citenamefont {Alipour}, \citenamefont {Benatti}, \citenamefont
  {Floreanini},\ and\ \citenamefont {Rezakhani}}]{marcantoni2017entropy}%
  \BibitemOpen
  \bibfield  {author} {\bibinfo {author} {\bibfnamefont {S.}~\bibnamefont
  {Marcantoni}}, \bibinfo {author} {\bibfnamefont {S.}~\bibnamefont {Alipour}},
  \bibinfo {author} {\bibfnamefont {F.}~\bibnamefont {Benatti}}, \bibinfo
  {author} {\bibfnamefont {R.}~\bibnamefont {Floreanini}}, \ and\ \bibinfo
  {author} {\bibfnamefont {A.~T.}\ \bibnamefont {Rezakhani}},\ }\bibfield
  {title} {\enquote {\bibinfo {title} {Entropy production and non-markovian
  dynamical maps},}\ }\href {\doibase 10.1038/s41598-017-12595-x} {\bibfield
  {journal} {\bibinfo  {journal} {Scientific Reports}\ }\textbf {\bibinfo
  {volume} {7}},\ \bibinfo {pages} {12447} (\bibinfo {year}
  {2017})}\BibitemShut {NoStop}%
\bibitem [{\citenamefont {Bhattacharya}\ \emph {et~al.}(2017)\citenamefont
  {Bhattacharya}, \citenamefont {Misra}, \citenamefont {Mukhopadhyay},\ and\
  \citenamefont {Pati}}]{PhysRevA.95.012122}%
  \BibitemOpen
  \bibfield  {author} {\bibinfo {author} {\bibfnamefont {Samyadeb}\
  \bibnamefont {Bhattacharya}}, \bibinfo {author} {\bibfnamefont {Avijit}\
  \bibnamefont {Misra}}, \bibinfo {author} {\bibfnamefont {Chiranjib}\
  \bibnamefont {Mukhopadhyay}}, \ and\ \bibinfo {author} {\bibfnamefont
  {Arun~Kumar}\ \bibnamefont {Pati}},\ }\bibfield  {title} {\enquote {\bibinfo
  {title} {Exact master equation for a spin interacting with a spin bath:
  Non-markovianity and negative entropy production rate},}\ }\href {\doibase
  10.1103/PhysRevA.95.012122} {\bibfield  {journal} {\bibinfo  {journal} {Phys.
  Rev. A}\ }\textbf {\bibinfo {volume} {95}},\ \bibinfo {pages} {012122}
  (\bibinfo {year} {2017})}\BibitemShut {NoStop}%
\bibitem [{\citenamefont {Popovic}\ \emph {et~al.}(2018)\citenamefont
  {Popovic}, \citenamefont {Vacchini},\ and\ \citenamefont
  {Campbell}}]{PhysRevA.98.012130}%
  \BibitemOpen
  \bibfield  {author} {\bibinfo {author} {\bibfnamefont {Maria}\ \bibnamefont
  {Popovic}}, \bibinfo {author} {\bibfnamefont {Bassano}\ \bibnamefont
  {Vacchini}}, \ and\ \bibinfo {author} {\bibfnamefont {Steve}\ \bibnamefont
  {Campbell}},\ }\bibfield  {title} {\enquote {\bibinfo {title} {Entropy
  production and correlations in a controlled non-markovian setting},}\ }\href
  {\doibase 10.1103/PhysRevA.98.012130} {\bibfield  {journal} {\bibinfo
  {journal} {Phys. Rev. A}\ }\textbf {\bibinfo {volume} {98}},\ \bibinfo
  {pages} {012130} (\bibinfo {year} {2018})}\BibitemShut {NoStop}%
\bibitem [{\citenamefont {Xu}\ \emph {et~al.}(2018)\citenamefont {Xu},
  \citenamefont {Liu},\ and\ \citenamefont {Feng}}]{PhysRevE.98.032102}%
  \BibitemOpen
  \bibfield  {author} {\bibinfo {author} {\bibfnamefont {Y.~Y.}\ \bibnamefont
  {Xu}}, \bibinfo {author} {\bibfnamefont {J.}~\bibnamefont {Liu}}, \ and\
  \bibinfo {author} {\bibfnamefont {M.}~\bibnamefont {Feng}},\ }\bibfield
  {title} {\enquote {\bibinfo {title} {Positive entropy production rate induced
  by non-markovianity},}\ }\href {\doibase 10.1103/PhysRevE.98.032102}
  {\bibfield  {journal} {\bibinfo  {journal} {Phys. Rev. E}\ }\textbf {\bibinfo
  {volume} {98}},\ \bibinfo {pages} {032102} (\bibinfo {year}
  {2018})}\BibitemShut {NoStop}%
\bibitem [{\citenamefont {Strasberg}\ and\ \citenamefont
  {Esposito}(2019)}]{PhysRevE.99.012120}%
  \BibitemOpen
  \bibfield  {author} {\bibinfo {author} {\bibfnamefont {Philipp}\ \bibnamefont
  {Strasberg}}\ and\ \bibinfo {author} {\bibfnamefont {Massimiliano}\
  \bibnamefont {Esposito}},\ }\bibfield  {title} {\enquote {\bibinfo {title}
  {Non-markovianity and negative entropy production rates},}\ }\href {\doibase
  10.1103/PhysRevE.99.012120} {\bibfield  {journal} {\bibinfo  {journal} {Phys.
  Rev. E}\ }\textbf {\bibinfo {volume} {99}},\ \bibinfo {pages} {012120}
  (\bibinfo {year} {2019})}\BibitemShut {NoStop}%
\bibitem [{\citenamefont {Rivas}(2020)}]{PhysRevLett.124.160601}%
  \BibitemOpen
  \bibfield  {author} {\bibinfo {author} {\bibfnamefont {\'Angel}\ \bibnamefont
  {Rivas}},\ }\bibfield  {title} {\enquote {\bibinfo {title} {Strong coupling
  thermodynamics of open quantum systems},}\ }\href {\doibase
  10.1103/PhysRevLett.124.160601} {\bibfield  {journal} {\bibinfo  {journal}
  {Phys. Rev. Lett.}\ }\textbf {\bibinfo {volume} {124}},\ \bibinfo {pages}
  {160601} (\bibinfo {year} {2020})}\BibitemShut {NoStop}%
\bibitem [{\citenamefont {de~Vega}\ and\ \citenamefont
  {Alonso}(2017)}]{RevModPhys.89.015001}%
  \BibitemOpen
  \bibfield  {author} {\bibinfo {author} {\bibfnamefont {In\'es}\ \bibnamefont
  {de~Vega}}\ and\ \bibinfo {author} {\bibfnamefont {Daniel}\ \bibnamefont
  {Alonso}},\ }\bibfield  {title} {\enquote {\bibinfo {title} {Dynamics of
  non-markovian open quantum systems},}\ }\href {\doibase
  10.1103/RevModPhys.89.015001} {\bibfield  {journal} {\bibinfo  {journal}
  {Rev. Mod. Phys.}\ }\textbf {\bibinfo {volume} {89}},\ \bibinfo {pages}
  {015001} (\bibinfo {year} {2017})}\BibitemShut {NoStop}%
\bibitem [{\citenamefont {Rivas}\ \emph {et~al.}(2010)\citenamefont {Rivas},
  \citenamefont {Plato}, \citenamefont {Huelga},\ and\ \citenamefont
  {Plenio}}]{Rivas_2010}%
  \BibitemOpen
  \bibfield  {author} {\bibinfo {author} {\bibfnamefont {{\'{A}}ngel}\
  \bibnamefont {Rivas}}, \bibinfo {author} {\bibfnamefont {A~Douglas~K}\
  \bibnamefont {Plato}}, \bibinfo {author} {\bibfnamefont {Susana~F}\
  \bibnamefont {Huelga}}, \ and\ \bibinfo {author} {\bibfnamefont {Martin~B}\
  \bibnamefont {Plenio}},\ }\bibfield  {title} {\enquote {\bibinfo {title}
  {Markovian master equations: a critical study},}\ }\href {\doibase
  10.1088/1367-2630/12/11/113032} {\bibfield  {journal} {\bibinfo  {journal}
  {New Journal of Physics}\ }\textbf {\bibinfo {volume} {12}},\ \bibinfo
  {pages} {113032} (\bibinfo {year} {2010})}\BibitemShut {NoStop}%
\bibitem [{\citenamefont {Gorini}\ \emph {et~al.}(1976)\citenamefont {Gorini},
  \citenamefont {Kossakowski},\ and\ \citenamefont
  {Sudarshan}}]{doi:10.1063/1.522979}%
  \BibitemOpen
  \bibfield  {author} {\bibinfo {author} {\bibfnamefont {Vittorio}\
  \bibnamefont {Gorini}}, \bibinfo {author} {\bibfnamefont {Andrzej}\
  \bibnamefont {Kossakowski}}, \ and\ \bibinfo {author} {\bibfnamefont
  {E.~C.~G.}\ \bibnamefont {Sudarshan}},\ }\bibfield  {title} {\enquote
  {\bibinfo {title} {Completely positive dynamical semigroups of n‐level
  systems},}\ }\href {\doibase 10.1063/1.522979} {\bibfield  {journal}
  {\bibinfo  {journal} {Journal of Mathematical Physics}\ }\textbf {\bibinfo
  {volume} {17}},\ \bibinfo {pages} {821--825} (\bibinfo {year}
  {1976})}\BibitemShut {NoStop}%
\bibitem [{\citenamefont {Lindblad}(1976)}]{lindblad1976generators}%
  \BibitemOpen
  \bibfield  {author} {\bibinfo {author} {\bibfnamefont {G.}~\bibnamefont
  {Lindblad}},\ }\bibfield  {title} {\enquote {\bibinfo {title} {On the
  generators of quantum dynamical semigroups},}\ }\href {\doibase
  10.1007/BF01608499} {\bibfield  {journal} {\bibinfo  {journal}
  {Communications in Mathematical Physics}\ }\textbf {\bibinfo {volume} {48}},\
  \bibinfo {pages} {119--130} (\bibinfo {year} {1976})}\BibitemShut {NoStop}%
\bibitem [{\citenamefont {Spohn}(1978)}]{doi:10.1063/1.523789}%
  \BibitemOpen
  \bibfield  {author} {\bibinfo {author} {\bibfnamefont {Herbert}\ \bibnamefont
  {Spohn}},\ }\bibfield  {title} {\enquote {\bibinfo {title} {Entropy
  production for quantum dynamical semigroups},}\ }\href {\doibase
  10.1063/1.523789} {\bibfield  {journal} {\bibinfo  {journal} {Journal of
  Mathematical Physics}\ }\textbf {\bibinfo {volume} {19}},\ \bibinfo {pages}
  {1227--1230} (\bibinfo {year} {1978})}\BibitemShut {NoStop}%
\bibitem [{\citenamefont {Wilde}(2013)}]{wilde2013quantum}%
  \BibitemOpen
  \bibfield  {author} {\bibinfo {author} {\bibfnamefont {Mark~M}\ \bibnamefont
  {Wilde}},\ }\href@noop {} {\emph {\bibinfo {title} {Quantum information
  theory}}}\ (\bibinfo  {publisher} {Cambridge University Press},\ \bibinfo
  {address} {Cambridge, England},\ \bibinfo {year} {2013})\BibitemShut
  {NoStop}%
\bibitem [{Note1()}]{Note1}%
  \BibitemOpen
  \bibinfo {note} {Throughout the present paper, we intentionally use the term
  thermodynamic entropy to distinguish it from other types of entropy, such as
  von Neumann entropy \cite {von2018mathematical}, R{\'e}nyi entropy \cite
  {renyi1961measures}, and a diagonal entropy \cite
  {POLKOVNIKOV2011486}}\BibitemShut {NoStop}%
\bibitem [{\citenamefont {D'Alessio}\ \emph {et~al.}(2016)\citenamefont
  {D'Alessio}, \citenamefont {Kafri}, \citenamefont {Polkovnikov},\ and\
  \citenamefont {Rigol}}]{doi:10.1080/00018732.2016.1198134}%
  \BibitemOpen
  \bibfield  {author} {\bibinfo {author} {\bibfnamefont {Luca}\ \bibnamefont
  {D'Alessio}}, \bibinfo {author} {\bibfnamefont {Yariv}\ \bibnamefont
  {Kafri}}, \bibinfo {author} {\bibfnamefont {Anatoli}\ \bibnamefont
  {Polkovnikov}}, \ and\ \bibinfo {author} {\bibfnamefont {Marcos}\
  \bibnamefont {Rigol}},\ }\bibfield  {title} {\enquote {\bibinfo {title} {From
  quantum chaos and eigenstate thermalization to statistical mechanics and
  thermodynamics},}\ }\href {\doibase 10.1080/00018732.2016.1198134} {\bibfield
   {journal} {\bibinfo  {journal} {Advances in Physics}\ }\textbf {\bibinfo
  {volume} {65}},\ \bibinfo {pages} {239--362} (\bibinfo {year}
  {2016})}\BibitemShut {NoStop}%
\bibitem [{\citenamefont {Anderson}(1958)}]{PhysRev.109.1492}%
  \BibitemOpen
  \bibfield  {author} {\bibinfo {author} {\bibfnamefont {P.~W.}\ \bibnamefont
  {Anderson}},\ }\bibfield  {title} {\enquote {\bibinfo {title} {Absence of
  diffusion in certain random lattices},}\ }\href {\doibase
  10.1103/PhysRev.109.1492} {\bibfield  {journal} {\bibinfo  {journal} {Phys.
  Rev.}\ }\textbf {\bibinfo {volume} {109}},\ \bibinfo {pages} {1492--1505}
  (\bibinfo {year} {1958})}\BibitemShut {NoStop}%
\bibitem [{\citenamefont {Anderson}(1961)}]{PhysRev.124.41}%
  \BibitemOpen
  \bibfield  {author} {\bibinfo {author} {\bibfnamefont {P.~W.}\ \bibnamefont
  {Anderson}},\ }\bibfield  {title} {\enquote {\bibinfo {title} {Localized
  magnetic states in metals},}\ }\href {\doibase 10.1103/PhysRev.124.41}
  {\bibfield  {journal} {\bibinfo  {journal} {Phys. Rev.}\ }\textbf {\bibinfo
  {volume} {124}},\ \bibinfo {pages} {41--53} (\bibinfo {year}
  {1961})}\BibitemShut {NoStop}%
\bibitem [{\citenamefont {Fano}(1961)}]{PhysRev.124.1866}%
  \BibitemOpen
  \bibfield  {author} {\bibinfo {author} {\bibfnamefont {U.}~\bibnamefont
  {Fano}},\ }\bibfield  {title} {\enquote {\bibinfo {title} {Effects of
  configuration interaction on intensities and phase shifts},}\ }\href
  {\doibase 10.1103/PhysRev.124.1866} {\bibfield  {journal} {\bibinfo
  {journal} {Phys. Rev.}\ }\textbf {\bibinfo {volume} {124}},\ \bibinfo {pages}
  {1866--1878} (\bibinfo {year} {1961})}\BibitemShut {NoStop}%
\bibitem [{\citenamefont
  {Friedrichs}(1948)}]{https://doi.org/10.1002/cpa.3160010404}%
  \BibitemOpen
  \bibfield  {author} {\bibinfo {author} {\bibfnamefont {K.~O.}\ \bibnamefont
  {Friedrichs}},\ }\bibfield  {title} {\enquote {\bibinfo {title} {On the
  perturbation of continuous spectra},}\ }\href {\doibase
  https://doi.org/10.1002/cpa.3160010404} {\bibfield  {journal} {\bibinfo
  {journal} {Communications on Pure and Applied Mathematics}\ }\textbf
  {\bibinfo {volume} {1}},\ \bibinfo {pages} {361--406} (\bibinfo {year}
  {1948})}\BibitemShut {NoStop}%
\bibitem [{\citenamefont {Lee}(1954)}]{PhysRev.95.1329}%
  \BibitemOpen
  \bibfield  {author} {\bibinfo {author} {\bibfnamefont {T.~D.}\ \bibnamefont
  {Lee}},\ }\bibfield  {title} {\enquote {\bibinfo {title} {Some special
  examples in renormalizable field theory},}\ }\href {\doibase
  10.1103/PhysRev.95.1329} {\bibfield  {journal} {\bibinfo  {journal} {Phys.
  Rev.}\ }\textbf {\bibinfo {volume} {95}},\ \bibinfo {pages} {1329--1334}
  (\bibinfo {year} {1954})}\BibitemShut {NoStop}%
\bibitem [{\citenamefont {Zhang}\ \emph {et~al.}(2015)\citenamefont {Zhang},
  \citenamefont {Lo}, \citenamefont {Xiong}, \citenamefont {Tu},\ and\
  \citenamefont {Nori}}]{PhysRevLett.115.168902}%
  \BibitemOpen
  \bibfield  {author} {\bibinfo {author} {\bibfnamefont {Wei-Min}\ \bibnamefont
  {Zhang}}, \bibinfo {author} {\bibfnamefont {Ping-Yuan}\ \bibnamefont {Lo}},
  \bibinfo {author} {\bibfnamefont {Heng-Na}\ \bibnamefont {Xiong}}, \bibinfo
  {author} {\bibfnamefont {Matisse Wei-Yuan}\ \bibnamefont {Tu}}, \ and\
  \bibinfo {author} {\bibfnamefont {Franco}\ \bibnamefont {Nori}},\ }\bibfield
  {title} {\enquote {\bibinfo {title} {Zhang et al. reply:},}\ }\href {\doibase
  10.1103/PhysRevLett.115.168902} {\bibfield  {journal} {\bibinfo  {journal}
  {Phys. Rev. Lett.}\ }\textbf {\bibinfo {volume} {115}},\ \bibinfo {pages}
  {168902} (\bibinfo {year} {2015})}\BibitemShut {NoStop}%
\bibitem [{\citenamefont {Caldeira}\ and\ \citenamefont
  {Leggett}(1983)}]{CALDEIRA1983587}%
  \BibitemOpen
  \bibfield  {author} {\bibinfo {author} {\bibfnamefont {A.O.}\ \bibnamefont
  {Caldeira}}\ and\ \bibinfo {author} {\bibfnamefont {A.J.}\ \bibnamefont
  {Leggett}},\ }\bibfield  {title} {\enquote {\bibinfo {title} {Path integral
  approach to quantum brownian motion},}\ }\href {\doibase
  https://doi.org/10.1016/0378-4371(83)90013-4} {\bibfield  {journal} {\bibinfo
   {journal} {Physica A: Statistical Mechanics and its Applications}\ }\textbf
  {\bibinfo {volume} {121}},\ \bibinfo {pages} {587--616} (\bibinfo {year}
  {1983})}\BibitemShut {NoStop}%
\bibitem [{\citenamefont {Serafini}(2017)}]{serafini2017quantum}%
  \BibitemOpen
  \bibfield  {author} {\bibinfo {author} {\bibfnamefont {Alessio}\ \bibnamefont
  {Serafini}},\ }\href@noop {} {\emph {\bibinfo {title} {Quantum Continuous
  Variables: A Primer of Theoretical Methods}}}\ (\bibinfo  {publisher} {CRC
  Press},\ \bibinfo {address} {Boca Raton, FL},\ \bibinfo {year}
  {2017})\BibitemShut {NoStop}%
\bibitem [{\citenamefont {Adesso}\ \emph {et~al.}(2014)\citenamefont {Adesso},
  \citenamefont {Ragy},\ and\ \citenamefont
  {Lee}}]{doi:10.1142/S1230161214400010}%
  \BibitemOpen
  \bibfield  {author} {\bibinfo {author} {\bibfnamefont {Gerardo}\ \bibnamefont
  {Adesso}}, \bibinfo {author} {\bibfnamefont {Sammy}\ \bibnamefont {Ragy}}, \
  and\ \bibinfo {author} {\bibfnamefont {Antony~R.}\ \bibnamefont {Lee}},\
  }\bibfield  {title} {\enquote {\bibinfo {title} {Continuous variable quantum
  information: Gaussian states and beyond},}\ }\href {\doibase
  10.1142/S1230161214400010} {\bibfield  {journal} {\bibinfo  {journal} {Open
  Systems \& Information Dynamics}\ }\textbf {\bibinfo {volume} {21}},\
  \bibinfo {pages} {1440001} (\bibinfo {year} {2014})}\BibitemShut {NoStop}%
\bibitem [{\citenamefont {Weedbrook}\ \emph {et~al.}(2012)\citenamefont
  {Weedbrook}, \citenamefont {Pirandola}, \citenamefont {Garc\'{\i}a-Patr\'on},
  \citenamefont {Cerf}, \citenamefont {Ralph}, \citenamefont {Shapiro},\ and\
  \citenamefont {Lloyd}}]{RevModPhys.84.621}%
  \BibitemOpen
  \bibfield  {author} {\bibinfo {author} {\bibfnamefont {Christian}\
  \bibnamefont {Weedbrook}}, \bibinfo {author} {\bibfnamefont {Stefano}\
  \bibnamefont {Pirandola}}, \bibinfo {author} {\bibfnamefont {Ra\'ul}\
  \bibnamefont {Garc\'{\i}a-Patr\'on}}, \bibinfo {author} {\bibfnamefont
  {Nicolas~J.}\ \bibnamefont {Cerf}}, \bibinfo {author} {\bibfnamefont
  {Timothy~C.}\ \bibnamefont {Ralph}}, \bibinfo {author} {\bibfnamefont
  {Jeffrey~H.}\ \bibnamefont {Shapiro}}, \ and\ \bibinfo {author}
  {\bibfnamefont {Seth}\ \bibnamefont {Lloyd}},\ }\bibfield  {title} {\enquote
  {\bibinfo {title} {Gaussian quantum information},}\ }\href {\doibase
  10.1103/RevModPhys.84.621} {\bibfield  {journal} {\bibinfo  {journal} {Rev.
  Mod. Phys.}\ }\textbf {\bibinfo {volume} {84}},\ \bibinfo {pages} {621--669}
  (\bibinfo {year} {2012})}\BibitemShut {NoStop}%
\bibitem [{\citenamefont {Wang}\ \emph {et~al.}(2007)\citenamefont {Wang},
  \citenamefont {Hiroshima}, \citenamefont {Tomita},\ and\ \citenamefont
  {Hayashi}}]{WANG20071}%
  \BibitemOpen
  \bibfield  {author} {\bibinfo {author} {\bibfnamefont {Xiang-Bin}\
  \bibnamefont {Wang}}, \bibinfo {author} {\bibfnamefont {Tohya}\ \bibnamefont
  {Hiroshima}}, \bibinfo {author} {\bibfnamefont {Akihisa}\ \bibnamefont
  {Tomita}}, \ and\ \bibinfo {author} {\bibfnamefont {Masahito}\ \bibnamefont
  {Hayashi}},\ }\bibfield  {title} {\enquote {\bibinfo {title} {Quantum
  information with gaussian states},}\ }\href {\doibase
  https://doi.org/10.1016/j.physrep.2007.04.005} {\bibfield  {journal}
  {\bibinfo  {journal} {Physics Reports}\ }\textbf {\bibinfo {volume} {448}},\
  \bibinfo {pages} {1--111} (\bibinfo {year} {2007})}\BibitemShut {NoStop}%
\bibitem [{\citenamefont {Ferraro}\ \emph {et~al.}(2005)\citenamefont
  {Ferraro}, \citenamefont {Olivares},\ and\ \citenamefont
  {Paris}}]{ferraro2005gaussian}%
  \BibitemOpen
  \bibfield  {author} {\bibinfo {author} {\bibfnamefont {Alessandro}\
  \bibnamefont {Ferraro}}, \bibinfo {author} {\bibfnamefont {Stefano}\
  \bibnamefont {Olivares}}, \ and\ \bibinfo {author} {\bibfnamefont
  {Matteo~GA}\ \bibnamefont {Paris}},\ }\href@noop {} {\emph {\bibinfo {title}
  {Gaussian States in Quantum Information}}}\ (\bibinfo  {publisher}
  {Biliopolis},\ \bibinfo {address} {Napoli},\ \bibinfo {year}
  {2005})\BibitemShut {NoStop}%
\bibitem [{\citenamefont {Banchi}\ \emph {et~al.}(2015)\citenamefont {Banchi},
  \citenamefont {Braunstein},\ and\ \citenamefont
  {Pirandola}}]{PhysRevLett.115.260501}%
  \BibitemOpen
  \bibfield  {author} {\bibinfo {author} {\bibfnamefont {Leonardo}\
  \bibnamefont {Banchi}}, \bibinfo {author} {\bibfnamefont {Samuel~L.}\
  \bibnamefont {Braunstein}}, \ and\ \bibinfo {author} {\bibfnamefont
  {Stefano}\ \bibnamefont {Pirandola}},\ }\bibfield  {title} {\enquote
  {\bibinfo {title} {Quantum fidelity for arbitrary gaussian states},}\ }\href
  {\doibase 10.1103/PhysRevLett.115.260501} {\bibfield  {journal} {\bibinfo
  {journal} {Phys. Rev. Lett.}\ }\textbf {\bibinfo {volume} {115}},\ \bibinfo
  {pages} {260501} (\bibinfo {year} {2015})}\BibitemShut {NoStop}%
\bibitem [{\citenamefont {Von~Neumann}(2018)}]{von2018mathematical}%
  \BibitemOpen
  \bibfield  {author} {\bibinfo {author} {\bibfnamefont {John}\ \bibnamefont
  {Von~Neumann}},\ }\href@noop {} {\emph {\bibinfo {title} {Mathematical
  foundations of quantum mechanics: New edition}}}\ (\bibinfo  {publisher}
  {Princeton university press},\ \bibinfo {address} {Princeton, NJ},\ \bibinfo
  {year} {2018})\BibitemShut {NoStop}%
\bibitem [{\citenamefont {R{\'e}nyi}(1961)}]{renyi1961measures}%
  \BibitemOpen
  \bibfield  {author} {\bibinfo {author} {\bibfnamefont {Alfr{\'e}d}\
  \bibnamefont {R{\'e}nyi}},\ }\enquote {\bibinfo {title} {On measures of
  entropy and information},}\ in\ \href@noop {} {\emph {\bibinfo {booktitle}
  {Proceedings of the Fourth Berkeley Symposium on Mathematical Statistics and
  Probability, Volume 1: Contributions to the Theory of Statistics}}}\
  (\bibinfo  {publisher} {University of California Press},\ \bibinfo {address}
  {Berkeley, California},\ \bibinfo {year} {1961})\ p.\ \bibinfo {pages}
  {547–561}\BibitemShut {NoStop}%
\bibitem [{\citenamefont {Polkovnikov}(2011)}]{POLKOVNIKOV2011486}%
  \BibitemOpen
  \bibfield  {author} {\bibinfo {author} {\bibfnamefont {Anatoli}\ \bibnamefont
  {Polkovnikov}},\ }\bibfield  {title} {\enquote {\bibinfo {title} {Microscopic
  diagonal entropy and its connection to basic thermodynamic relations},}\
  }\href {\doibase https://doi.org/10.1016/j.aop.2010.08.004} {\bibfield
  {journal} {\bibinfo  {journal} {Annals of Physics}\ }\textbf {\bibinfo
  {volume} {326}},\ \bibinfo {pages} {486--499} (\bibinfo {year}
  {2011})}\BibitemShut {NoStop}%
\end{thebibliography}%
\end{document}